\documentclass[smallextended]{svjour3}       
\smartqed  
\usepackage{graphicx}
\usepackage{amsmath}
\usepackage[square,sort,comma,numbers]{natbib}
\usepackage{pdfpages}

\def\aap{A\&A}

\begin{document}

\title{Why Chromatic Imaging Matters}

\author{Joel Sanchez-Bermudez \and
	Florentin Millour \and Fabien Baron \and Roy van Boekel \and Laurent Bourg\`es \and Gilles Duvert  \and Paulo J. V. Garcia \and Nuno Gomes \and Karl-Heinz Hofmann  \and Thomas Henning \and Jacob W. Isbell \and Bruno Lopez \and Alexis Matter  \and J-Uwe Pott \and Dieter Schertl \and Eric Thi\'ebaut \and Gerd Weigelt \and John Young 
       } 

\authorrunning{J. Sanchez et al.}

\institute{
            J. Sanchez-Bermudez works at \at European Southern Observatory, Alonso de C\'ordova 3107, Vitacura, Santiago de Chile \email{sanchezj@eso.org}
\and B. Lopez, F. Millour, and A. Matter work at \at Universit\'{e} C\^{o}te d'Azur, OCA, CNRS, Lagrange, Parc Valrose, B\^{a}t. Fizeau, 06108 Nice cedex 02, France \email{fmillour@oca.eu} 
\and F. Baron works at \at Department of Physics and Astronomy, Georgia State University P.O. Box 5060, Atlanta, GA 30302-5060, USA
\and G. Duvert and L. Bourg\`es work at \at Univ. Grenoble Alpes, CNRS, IPAG, F-38000 Grenoble, France
\and K-H Hofmann, D. Schertl, and G. Weigelt work at \at Max-Planck-Institut f\"ur Radioastronomie, Auf dem H\"ugel 69, Bonn 53121, Germany 
\and P. J. V. Garcia works at \at Universidade do Porto - Faculdade de Engenharia, Departamento de Engenharia F\'isica -CENTRA-, Porto, Portugal 
\and N. Gomes works at \at NUCLIO - N\'ucleo Interativo de Astronomia, Largo dos Top\'azios, 48, 3$^{o}$ Frente, 2785-817 S\~ao Domingos de Rana, Portugal
\and T. Henning, J-Uwe Pott, R. van Boekel, and J.W. Isbell work at \at Max-Planck-Institut f\"ur Astronomie, K\"onigstuhl 17, Heidelberg 69117, Germany 
\and E. Thi\'ebaut works at \at Centre de Recherche Astrophysique de Lyon, Lyon, France 
\and J. Young works at \at University of Cambridge, Cambridge, United Kingdom.
}
\date{Received: date / Accepted: date}

\maketitle

\begin{abstract}
During the last two decades, the first generation of beam combiners at the Very Large Telescope
Interferometer has proved the importance of optical interferometry for high-angular resolution
astrophysical studies in the near- and mid-infrared. With the advent of 4-beam combiners at the
VLTI, the $u-v$ coverage per pointing increases significantly, providing an opportunity to use
reconstructed images as powerful scientific tools. Therefore, interferometric imaging is already a key feature of the new generation of VLTI instruments, as well as for other interferometric facilities like CHARA and JWST. It is thus imperative to account for the current image reconstruction capabilities and their expected evolutions in the coming years. Here, we present a general overview of the current situation of optical interferometric image reconstruction with a focus on new wavelength-dependent information, highlighting its main advantages and limitations. As an appendix we include several cookbooks describing the usage and installation of several state-of-the art image reconstruction packages. To illustrate the current capabilities of the software available to the community, we recovered chromatic images, from simulated MATISSE data, using the MCMC software SQUEEZE. With these images, we aim at showing the importance of selecting good regularization functions and their impact on the reconstruction. 
\end{abstract}

\section{Introduction}
\label{intro}

Except perhaps for the most simple objects, interferometric data are hard to
interpret directly, and image reconstruction is a powerful tool for scientific analysis of the observations. However, optical long-baseline interferometric instruments have, since their beginnings, faced a series of challenges to produce science-grade images. These challenges are related to:

\begin{itemize}
\item Sparse sampling of the measurements due to a limited number
of spatial frequencies;
\item Non-convex inverse problem to solve (i.e., there may be several local minima);
\item Phase disturbance by the atmosphere in front of the telescope or interferometer, which smears out the phase information of the object.
\end{itemize} 

Since direct inversion of the data is neither possible nor recommended, image
reconstruction algorithms are based on regularized minimization processes that iteratively solve an (inverse) \emph{ill-posed
problem}, where the model of the data given the image is compared to the actual
data, in order to determine how to better fit the data while
respecting some imposed constraints.  These
constraints are needed to avoid over-fitting of the data (and thus explaining, erroneously,
the noise as well as the significant signal) and to compensate for the sparsity
of the data which leads to an under-determined problem. The basic concept of this regularized minimization could be expressed in the following form:

\begin{equation}
\boldsymbol{x}_{\mathrm{ML}}=\underset{\boldsymbol{x}}{\textrm{argmin}}[(1/2) \chi^2(\boldsymbol{x})+\sum_{i}^{n}\mu_i R(\boldsymbol{x})_i]\,,\\
\end{equation}
where $\chi^2(\boldsymbol{x})$ is the likelihood of our data to a
given imaging model,  $R(\boldsymbol{x})_i$ are the used (prior) regularization functions and $\mu_i$ the weighting factors that trade-off
between the likelihood and priors. There are several algorithms to perform the minimization to find the most probable image $\boldsymbol{x}_{\mathrm{ML}}$. Two of the most important algorithms are Gradient Descent and Monte Carlo Markov Chain (MCMC). Gradient Descent is an optimization algorithm that takes steps proportional to the negative of a gradient function with respect to the image pixels in order to find the best solution. This method is fast; however, it may fall on local minima that could lead to a misleading solution in the image construction process. As a consequence, several types of gradient descent algorithms have been developed to improve the basic one, such as the Active Set Algorithm for Conjugate Gradient Descent (ASA-CG) \citep{Hager2005}.

On the other hand, MCMC is based on a random process which determines the flux element positions in a pixel grid until the desired distribution of pixel flux fits the data and reaches an equilibrium distribution. The great advantage of this method is that it could find a global minimum, at the cost of being significantly slower than Gradient Descent.

By using these methods, several image reconstruction software have been developed to
analyse optical/infrared interferometric data,
and most of them are available to the community. However, due to the nature of the minimization process, these software cannot be considered ``black boxes'' which could produce
science-grade images without any user interaction. 
The choice of regularization functions must be carried out with great attention \citep{Renard2011}, always based on the prior knowledge of the source's brightness distribution. 
This choice has a significant influence on the resulting image (as we explore later), with some regularization functions emphasizing image smoothness while others emphasize sharp edges or compactness of the brightness distribution. 
A good image reconstruction is therefore not just the matter of having a good algorithm but
also of using it correctly. 

While most software can be understood from the principles summarized above,
they still differ in a number of points. One of the most obvious difference is
the kind of regularizations implemented.  Some offer many
different possibilities, while others are more restricted. Another less
obvious difference, but not less important, is the likelihood term. This is mainly because optical interferometers do not provide complex visibilities
but other observables such as the powerspectrum, the closure phase and the
chromatic differential phase.  These quantities are less sensitive to random delays
induced by the atmosphere turbulence while the complex visibilities are completely
destroyed by them. At this time, there is no consensus on the statistics of
these observables and the image reconstruction algorithms implement different
expressions for the likelihood term.

During the last twenty years several algorithms have reached sufficient maturity to produce science-grade images, among them we have: the Building-Block Method \citep{Hofmann_1993}, BSMEM \citep{Buscher_1994}, MiRA \citep{Thiebaut_2008}, SQUEEZE \citep{Baron_2010b}, WISARD \citep{Meimon_2005}, and IRBis \citep{Hofmann_2014}. The capabilities of these software to recover images from optical/infrared long-baseline interferometric data have been shown several times in the community through the ``Interferometric Imaging Beauty Contest'' \citep[see e.g.,][]{Sanchez-Bermudez_2016c}. Additionally, these software have been used repeatedly to recover images from Fizeau interferometric data in Sparse Aperture Masking observations. For example, \citep{Tuthill_1999} used MACIM to recover the first images of Pinwheel nebulae; while \citep{Sanchez-Bermudez2014} used BSMEM to recover images of several bow shocks around WR-stars in the central parsec of our Galaxy. 

However, the standard use of several of these software packages was limited to recover monochromatic images. With the consolidation of the spectro-interferometric capabilities of the second generation instruments, GRAVITY \citep{Eisenhauer_2017} and MATISSE \citep{Lopez_2008}, at the Very Large Telescope Interferometer (VLTI), some of the aforementioned software have been updated, and new methods are being developed to obtain chromatic images in a systematic way. For example, at the VIII edition of the ``Interferometric Imaging Beauty Contest" a benchmark of different algorithms was done with simulated \textit{chromatic} data for the first time \citep{Sanchez-Bermudez_2016b}.

\section{The necessity of  chromatic imaging}
\label{sec:whychromaimag}

\subsection{Scientific cases for optical/infrared interferometry}
\label{sec:scicases}
 
Optical interferometry allows one to probe the innermost regions of several astrophysical environments. Here we describe three examples, but there are of course many more. Around young stars, where the accretion on the star itself is almost done and the gas and dust disk have settled, the process of planet formation occurs with planet accretion, planets migration, gap formation, etc. These regions measure only a few astronomical units across ($\sim 10$ au), i.e. for a system at 100\,pc, this corresponds to a hundred milli-arcseconds, i.e. barely the angular resolution of an 8\,m-class telescope. Therefore, the high angular resolution of long baseline interferometry is the only one that provide us the possibility to explore the inner regions (at scales of $\sim$ 1-10 milli-arcseconds) of such environments. However, such environments are quite complex and properly determining their structure is only possible if image reconstruction techniques are robust.

Another example is the study of the line-emitting regions in massive stars. Here, the photosphere of the star is not directly imaged to probe its nature and evolution, but the distribution and kinematics of the surrounding gas. These regions are also located at only few au from the star, but typically the sources are located at distances larger than 1kpc, preventing them from being resolved even with adaptive optics on 8m-class telescopes. While continuum optical interferometric images have been reconstructed for a few of these objects, studies of the line-emitting region have remained difficult. Chromatic imaging, however, was recently attempted for Phi Per and Eta Car \citep{Mourard_2015, Sanchez-Bermudez_2018}, where it was shown that the inclusion of differential phase information can vastly improve the image quality.

The third example is asymptotic giant branch (AGB) stars. These objects are one of the final steps in the evolution of low- and intermediate-mass stars. Since they return most of their processed material to the interstellar medium, they are important for understanding the chemical evolution of galaxies \citep{Herwig_2005}. Therefore, understanding the atmospheric structure is crucial to bridge the physics of the enriched stellar interior and the evolution-driven winds. Chromatic interferometric imaging allows us to resolve the structure of the photosphere and together with the inner dust-shell layers, allowing us to investigate convection and the formation of dust at the most compact scales, among other physical phenomena \citep[see e.g.,][]{Ohnaka_2017, Paladini_2018}.

\subsection{Solving the phase problem}
\label{sec:phaseprob}

The problem of image reconstruction was faced already a few decades ago in radio astronomy. Several methods like CLEAN and the Maximum Entropy Method (MEM) were developed to produce images out of visibilities and phases. However, it was not until a new set of techniques were developed (e.g., hybrid mapping and self-calibration) that imaging with radio telescopes came to a boom, allowing the recovery of spectral images \citep[see e.g.,][]{Pearson_1984, Readhead_1988}. In optical/infrared long-baseline interferometry, it was considered up to very recently that the phase information was completely lost due to the turbulent atmosphere. Therefore, the powerspectrum (often called squared visibility) and the argument of the bispectrum (the so-called closure phase) were defined as the main observables of the geometrical information of the source's brightness distribution. However,  the ``new'' available wavelength-dependent phase (the so-called ``differential phase''), accessible through the spectro-interferometric capabilities of instruments like AMBER \citep{AMBER_Petrov_2007}, GRAVITY \citep{Eisenhauer_2017}, and MATISSE \citep{Lopez_2008} provide much more information than the geometric information of the closure phase alone.

The differential phase is an interferometric observable that provides information about the flux centroid position of the source's brightness distribution across the observed bandpass. This observable was not considered in optical/infrared interferometry imaging until \citep{Millour_2006, Schmitt_2009}, and \citep{Millour_2011}. Indeed, differential phase provides a corrugated phase measurement, which, in theory, can be incorporated into a self-calibration algorithm, in a very similar way to what is done in radio interferometry \citep{Pearson_1984}.

As early as 2003, J. Monnier anticipated ``revived activity [on self-calibration] as more interferometers with imaging capability begin to produce data''. And indeed, the conceptual bases for using differential phases in image reconstruction were laid in \citep{Millour_2006}. The Schmitt paper \citep{Schmitt_2009} was a first attempt to use differential phases in image reconstruction. They considered that the phase in the continuum was equal to zero, making it possible to use the differential phase (then equal to the phase) in the H$\alpha$ emission line of the $\beta$ Lyr system. In this way, they were able to image the shock region between the two stars at different orbital phases.

Millour et al. 2011\citep{Millour_2011} went one step further, by using an iterative process similar to radio-interferometry self-calibration, in order to reconstruct the phase of the object from the closure phases and differential phases. This way, they could recover the image of a rotating gas+dust disc around a supergiant star, whose image is asymmetric even in the continuum (non-zero phase). This method was subsequently used in a few papers to reconstruct images of supergiant stars \citep[see e.g.,][]{Ohnaka_2011, Ohnaka_2013} and Eta Carinae \citep[e.g.,][]{Weigelt_2016, Sanchez-Bermudez_2018}. More recently,\citep{Hone2017} published the first chromatic reconstruction of the Br$\gamma$ structure in a YSO with IRBis. \citep{Mourard_2015}
presented an extended method to tackle the image reconstruction challenges posed by the lacking of the closure phases and a proper calibration of the spectrally-dispersed visibilities. 

\subsection{Going one step further}
\label{sec:onemorestep}

An enormous effort has been made in the community to develop reconstruction methods one step further. For example, a project  named POLCA funded by the French ANR (Agence Nationale de la Recherche) made the following advances on the subject:

\begin{itemize}
\item Statistical analysis of AMBER data showed that the interferometric data do not follow the usual assumptions of uncorrelation and Gaussian (Normal) noise distribution. Correlations over time are significant and can be partly disentangled by considering differential visibilities in addition to absolute visibilities. A Student distribution of noise on the visibilities should be used instead of a Normal one it is expected since visibility is calculated as the division of two random variables, like in \citep{AMBER_Tatulli_2007}. This could lead to a future improvement on descent algorithms used in model-fitting or image reconstruction \citep[][]{Schutz_2014}.
\item New development on the core image reconstruction algorithm to take into account the wavelength-dependence of the data has been achieved and is distributed under the ``PAINTER'' software \citep{Schutz_2014}. It works on chromatic datasets (i.e. both the bispectrum and the differential phase) and produces chromatic image cubes by using the ADMM descent algorithm and spatio-spectral regularizations. An improved version, presented in \citep{Schutz_2015}, uses wavelets for spatial regularization and Discrete Cosine Transform (DCT) for spectral regularization.
\item A future chromatic image reconstruction algorithm is in development by F. Soulez \& E. Thi\'ebaut under the name ``MIRA-3D'' \citep{Soulez_2014}. This algorithm uses a joint spectro-spatial regularisation, and thus will be applicable to objects with components exhibiting different spectra.
\item The power of combining chromatic model-fitting and ``grey'' image reconstruction was demonstrated by \citep{Kluska_2014} for low-spectral resolution datasets. The software
``SPARCO'' was developed to demonstrate this. The potential of this technique is great and allows one to perform ``numerical coronagraphy'' on the interferometric data, by removing
the (main) contribution from the central star.
\end{itemize}

On the other hand, Fabien Baron, at Georgia State University, has developed a chromatic imaging algorithm called SQUEEZE \citep{Baron_2010, Baron_2010b, Baron_2012}. This software uses a MCMC as engine for the minimization. The code is able to use Simulated Annealing and Parallel Tempering with Metropolis-Hasting moves. SQUEEZE allows us to use (i) visibilities (amplitudes and phases), (ii) powerspectra, (iii) bispectra, or combination of them as input observables. One of the key characteristics of this software is the large number of regularization functions offered, like the L0-norm, the L2-norm, Entropy, Total Variation, Wavelets, etc. 

All these software efforts have had a tremendous impact in the development of chromatic imaging, and they could lead to a new generation of image reconstruction software. For example, combining the new core algorithms with the chromatic model-fitting features of ``SPARCO'' and self-calibration techniques could produce a leading edge image reconstruction suite suitable to reduce both archival AMBER data, and the new data being taken with the new spectro-imaging interferometric instruments GRAVITY, and MATISSE at the VLTI or MYSTIC \citep{ten-Brummelaar_2016a, ten-Brummelaar_2016b} at CHARA.

\subsection{Toward integrated tools for image reconstruction}
\label{sec:future}

With the developments mentioned above, one can produce images, not only by using the closure phases but also the wavelength-differential phases. These developments have allowed us to recover simultaneously chromatic images, while, additional techniques like self-calibration have allowed us to perform imaging even with 2 telescopes. Nowadays, the number of image reconstruction packages available to the community is large, however each of these packages uses different interfaces and programming languages, making a comparison relatively difficult in practice.

In this respect, there is a working group at JRA4 aiming at providing advances in easing the use of image reconstruction software through a web-based interface called OImaging\footnote{Available at http://www.jmmc.fr/oimaging part of the European Commission's FP7 Capacities programme (Grant Agreement Number 312430 )}, in a similar way to what has been done for model-fitting and the LITpro\footnote{LITpro software available at http://www.jmmc.fr/litpro} software at JMMC. Similarly the input/output formats (\texttt{OIFITS v1} vs \texttt{OIFITS v2}) and visualization tools should be standardized to allow the comparison of different software and different runs in the huge image-reconstruction parameter space. 

Also of interest is a series of complementary recipes to produce ``science-grade'' images like Low Frequency Filling, Monte Carlo analysis \citep{Millour_2012}, and other recipes (which have no specific name) like using central symmetry properties of the object \citep{Le_Bouquin_2009}, or reducing the field-of-view of the reconstruction to the photosphere of the central star only \citep{Monnier_2007}. The integration of such recipes with the available image reconstruction software, either hard-coding them in the software or with external tools, would be valuable for the coming generation of imaging software.

\section{An example application: the future MATISSE image reconstruction capacity, illustrated}
\label{sec:methods}

Here, we have highlighted the importance of image reconstruction to scientifically assess
the information encoded in the optical interferometry data. Therefore, characterizing the imaging
capabilities of the different interferometric arrays is necessary, especially in the frame of the
upcoming infrared beam-combiners. MATISSE (Multi-Aperture mid-Infrared SpectroScopic Experiment; \citep{Lopez_2008, Lopez_2009}) is one of the second-generation interferometric instruments of the VLTI.
This instrument is conceived to interconnect up to four telescopes, either the Unit Telescopes (UTs) or the Auxiliary Telescopes (ATs) to capture visibilities, closure phases and differential phases in the mid-infrared. It represents a major advance compared with its predecessor MIDI, mainly, because it will allow us to recover, for the first time, the closure phase at three different bands: L-M (2.8--5.2 $\mu$m), and N (8--13 $\mu$m).

\subsection{Data Simulation}

\begin{figure}
  \centering
    \includegraphics[width=\textwidth]{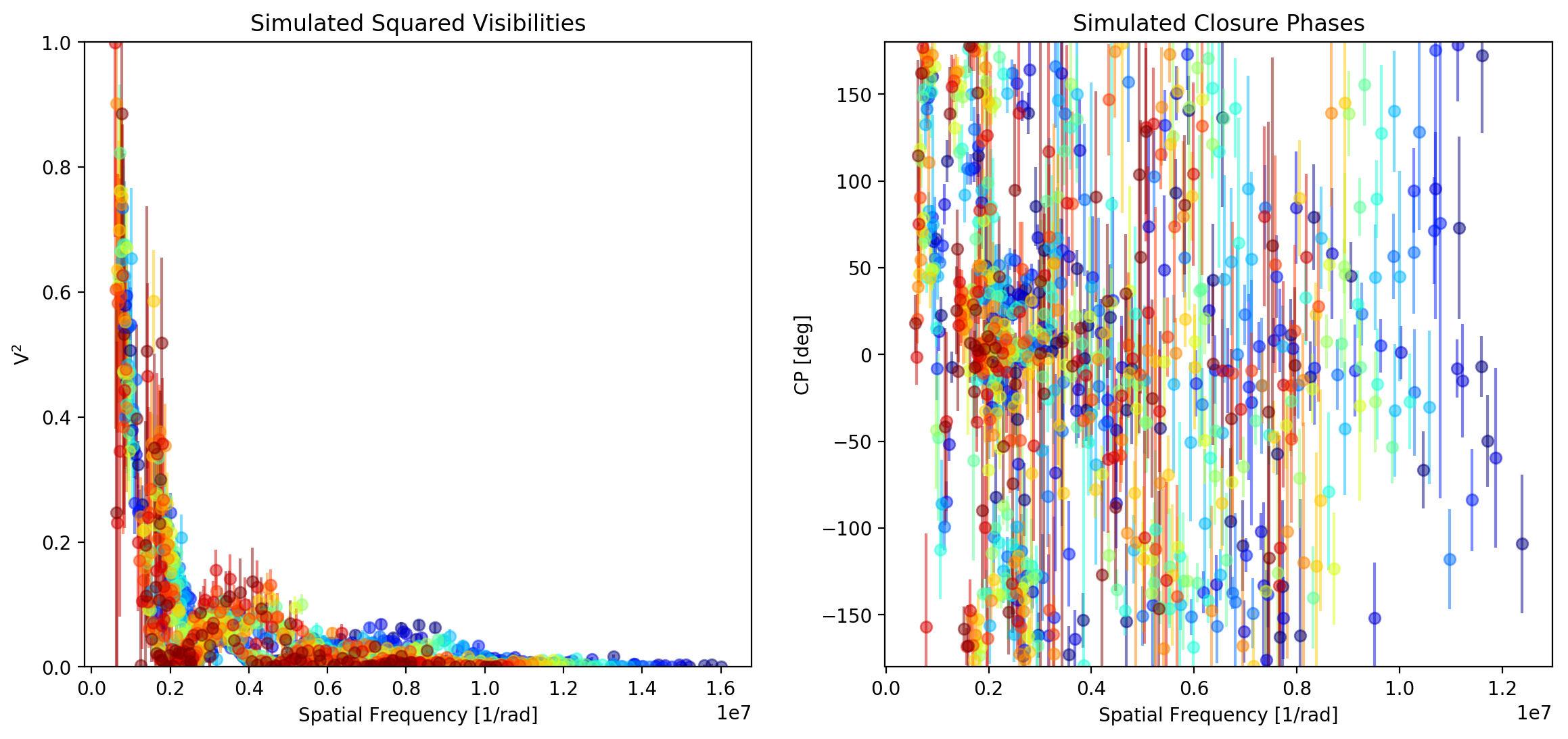}
  \caption{The figure displays the simulated squared visibilities (\textit{left}) and closure phases (\textit{right}) used to demonstrate the MATISSE imaging capabilities. The plotted observables correspond to three different observing nights, each one of them with a different ATs quadruplet. The colors represent the effective wavelength from 8$\mu$m (blue) to 13$\mu$m(red). Notice that the target is completely resolved and that it has a clear deviation from point-symmetry since the closure phases vary between -180$^{\circ}$ and 180$^{\circ}$.}
\label{fig:observables}
\end{figure}

One of the major science case studies of MATISSE is the characterization of proto-planetary discs
around young stellar objects. In this respect, image reconstruction represents a unique tool to obtain
constraints on (i) the physics in the inner discs, (ii) the signatures of
interaction between forming planets and the dusty disc, (iii) detection of companions in the disc-like
structure, (iv) the signatures tracing different dust mineralogy (e.g., the silicate feature at 10 $\mu$m) and
(v) the gas disc kinematics, among others.
Therefore, we selected a prototypical Herbig Ae star as our image reconstruction source. HD\,179\,218
is a B9 star with an effective temperature $T_{\mathrm{eff}}$ = 9600 $K$ \citep{Folsom_2012}, a stellar mass $M$ = 2.9
$M_{\odot}$ \citep{vanBoekel_2005} and located at a distance of 250 pc \citep{vanLeeuwen_2007}. Thanks
to a large collection of MIDI and Spectral Energy Distribution (SED) data, Menu et al. (in prep.) inferred a
disc with an eccentric inner gap. Nevertheless no image of this structure has been obtained so far. A
set of radiative transfer images obtained from the Menu et al. model was used to simulate the
expected MATISSE $u-v$ coverage. We simulated three different interferometric arrays with the ATs,
assuming MATISSE observations in low-resolution mode (R$\sim$35). The three simulated configurations
sample some of the small (A0-B2-C1-D0), medium (D0-G2-J3-K0) and large (A0-G1-J2-J3) telescope
configurations available at the VLTI. We considered that the target was successfully observed at six
different position angles over three different nights, each one with a different AT configuration.
The applied noise model was generated using the MATISSE simulator developed at the Observatoire
de la C\^{o}te d'Azur by Alexis Matter. This simulator uses the pre-computed theoretical interferometric observables and
adds two main types of noise: (i) the fundamental noise and (ii) the calibration noise. Once the
different error contributions are calculated, the theoretical observables are randomly changed
following a Gaussian distribution within the computed error-bars. Figure \ref{fig:observables} displays an example of
the squared visibilities and closure phases recovered for the simulated AT configurations. 

\subsection{Recovering the Images}

\begin{figure}
  \centering
    \includegraphics[width=0.8\textwidth]{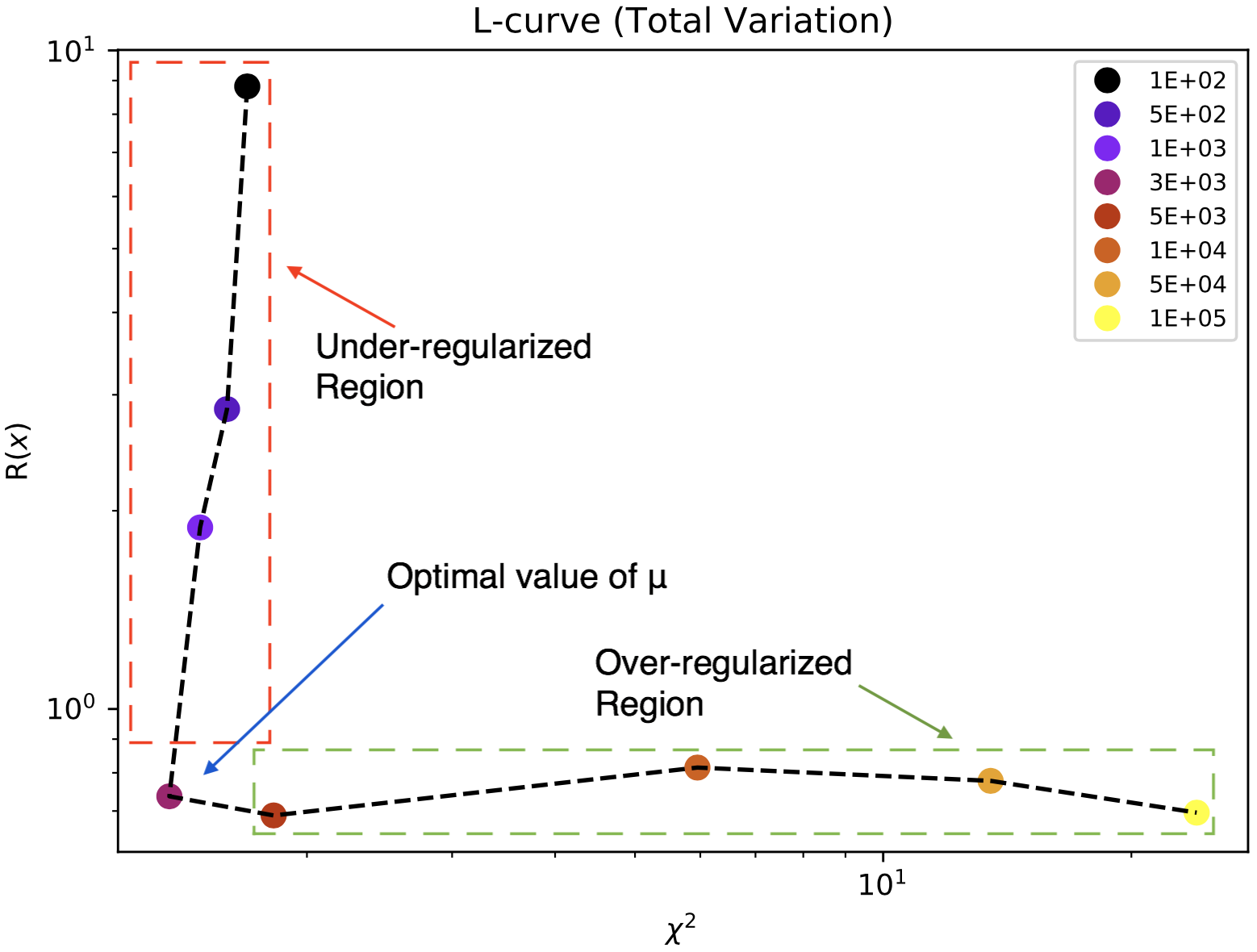}
  \caption{L-curve obtained with different values of $\mu$ using the Total Variation as regularizer. The vertical axis displays the value of $R(\boldsymbol{x})_i$ and the horizontal axis the $\chi^2(\boldsymbol{x})$. The optimal value of $\mu$ is at the elbow of the curve with the smallest $\chi^2$, in this case it is equal to 3.0e3. The optimal value of $\mu$, together with the under-regularized and over-regularized regions are labelled on the frame.}
\label{fig:lcurve_vals}
\end{figure}

\begin{figure}
  \centering
    \includegraphics[width=\textwidth]{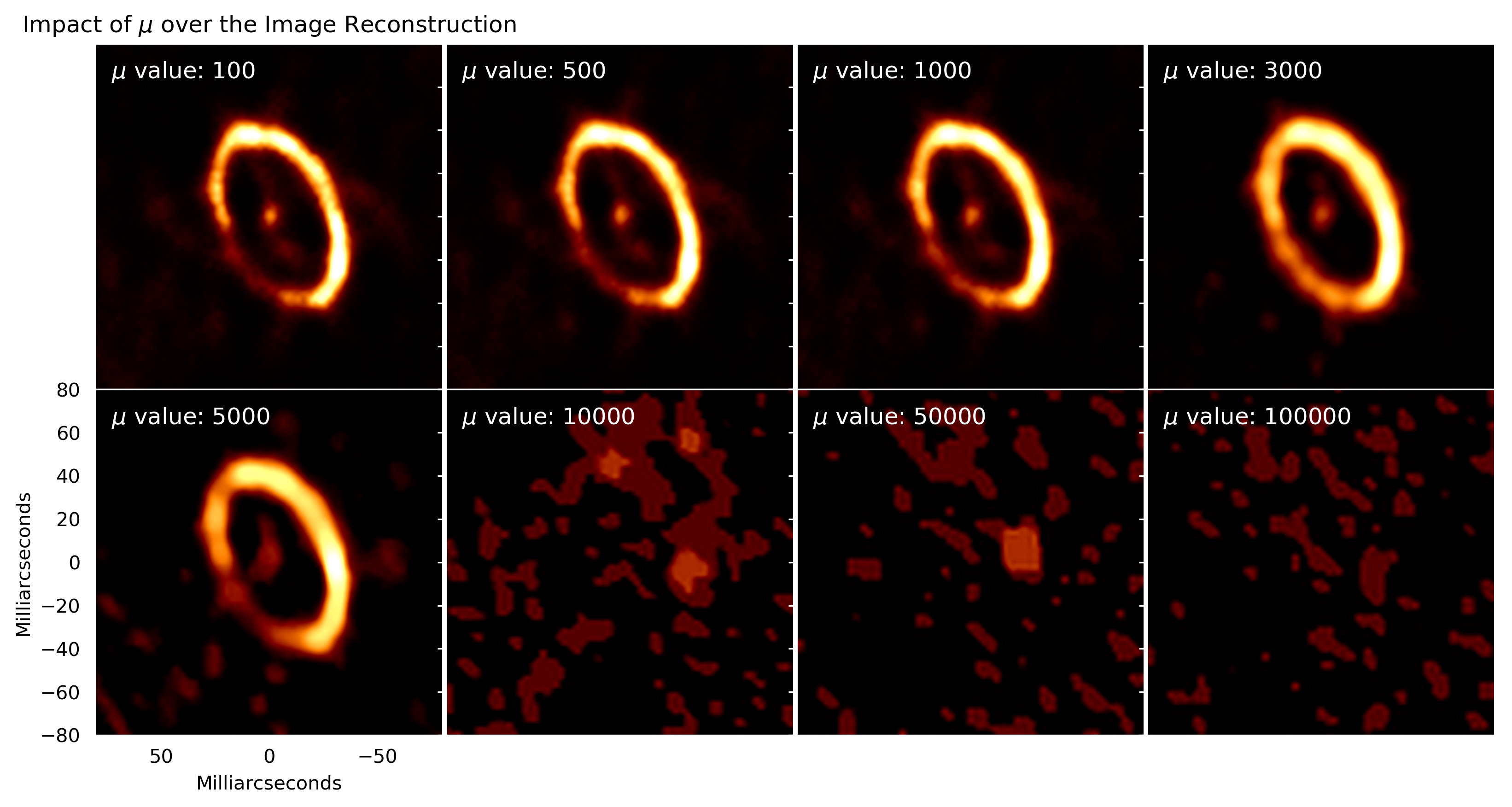}
  \caption{The image shows the impact of $\mu$ on the image reconstruction process. Eight reconstructed images of the source are shown using the hyperparameter values shown in Figure \ref{fig:lcurve_vals}. Notice that for the first three images (in the under-regularizer region) the general morphology of the object is recovered, but the reconstructed maps are clearly noisier than the one with the optimal value of $\mu = 3000$. However, when the value of the hyperparameter is quite large, the structure of the source is completely erased and the code is not able to converge to reproduce the expected brightness distribution of the source.}
\label{fig:ims_reg}
\end{figure}

To recover the images from our simulated data set, we used SQUEEZE. One of the key parameters in interferometric image reconstruction is the selection of the regularization functions. An important advantage of SQUEEZE is that it can process multiple regularization functions simultaneously, and because it uses MCMC sampling, they need not have well-defined gradients. The second important parameter for the
reconstruction is the hyperparameter $\mu$ that controls the trade-off between the likelihood and the prior
information of the brightness distribution encoded in the regularizers. Therefore, selecting the appropriate
value of $\mu$ is crucial for the image reconstruction process. One of the most common methods to
select the optimum $\mu$ is the L-curve \citep{Hansen_1992, Bose_2001, Kluska_2014, 2014A&A...569A..10D}. It computes the image
solution for several values of $\mu$, characterizing the response of the prior term versus the $\chi^2$. The
optimal values are normally at the first inflection point of the L-curve. Figure \ref{fig:lcurve_vals} displays an example of this method, where $R(\boldsymbol{x})_i$ (in this case the Total Variation) is compared with $\chi^2(\boldsymbol{x})$ for different values of $\mu$. The plot shows clearly three areas; the so-called ``under-regularized'' region, in which different values of the regularizer produce similar $\chi^2$; and inflection point, which corresponds to the optimal value of $\mu$; and the so-called ``over-regularized'' region, where the $\chi^2$ increases proportionally to each increment in the value of $\mu$. This occurs because the regularizer is dominating the convergence criteria of the algorithm. To illustrate these effects over the reconstruction, we display the role of $\mu$ over the reconstruction of our object in Figure \ref{fig:ims_reg}. 

\begin{figure}
  \centering
    \includegraphics[width=\textwidth]{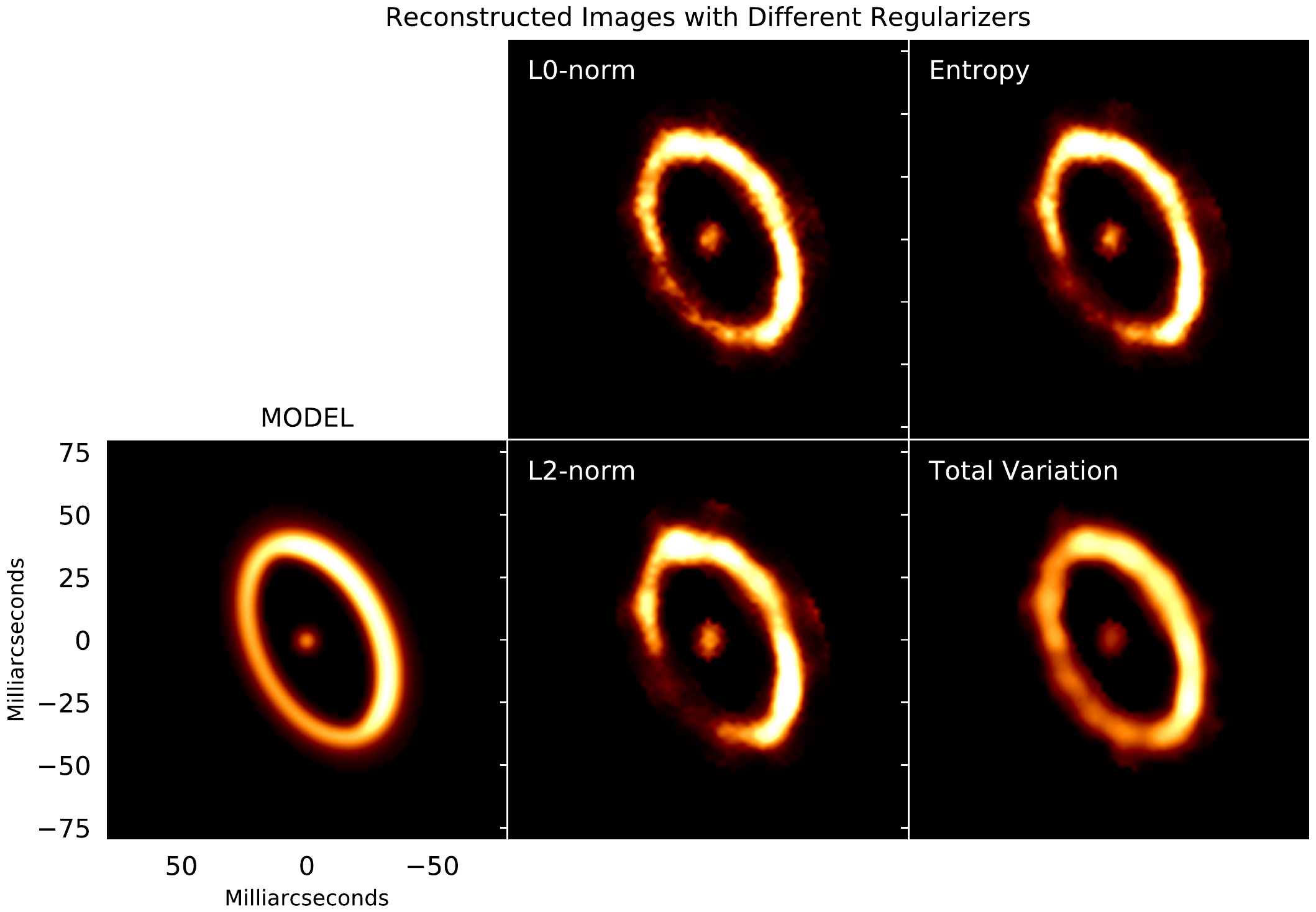}
  \caption{Best reconstructed monochromatic images using SQUEEZE with four different regularizers. The leftmost panel displays an average of the model image for comparison. The colormap of the images is normalized to the maximum pixel value in the model frame. Notice the dependence of the reconstructed image according to the used regularization function.}
\label{fig:monochromatic}
\end{figure}

Figure \ref{fig:monochromatic} shows the N-band monochromatic reconstruction obtained with four
different SQUEEZE regularization functions, to evaluate their impact. The target was considered to evolve as a grey-body object over all the channels inside the bandpass. Although this consideration is not true for most of the astrophysical objects, it represents a good
starting point to calibrate the different parameters used for the reconstruction. In this case, we
used both the squared visibilities and closure phases to recover the brightness distribution of
our object. From this reconstruction, we could notice that the different regularizers were able to
recover the general morphology of the target. Nevertheless, there were still some significant
differences among them. For example, while the Total Variation was able to recover a smooth rim
morphology, the other regularizers underestimated the brightness distribution of the rim for
position angles between 90$^{\circ}$ and 180$^{\circ}$ (East of North). Additionally, all the reconstructed images
show several bright-spots well localized along the rim, instead of a uniform distribution like in
the model. This highlights the importance of a good selection of the regularization function and demonstrates the necessity of user interaction with the reconstruction process. It also shows that the reconstructed images are still models of the object's brightness distribution and caution must be taken when physical parameters are derived from them. 

\begin{figure}
  \centering
    \includegraphics[width=\textwidth]{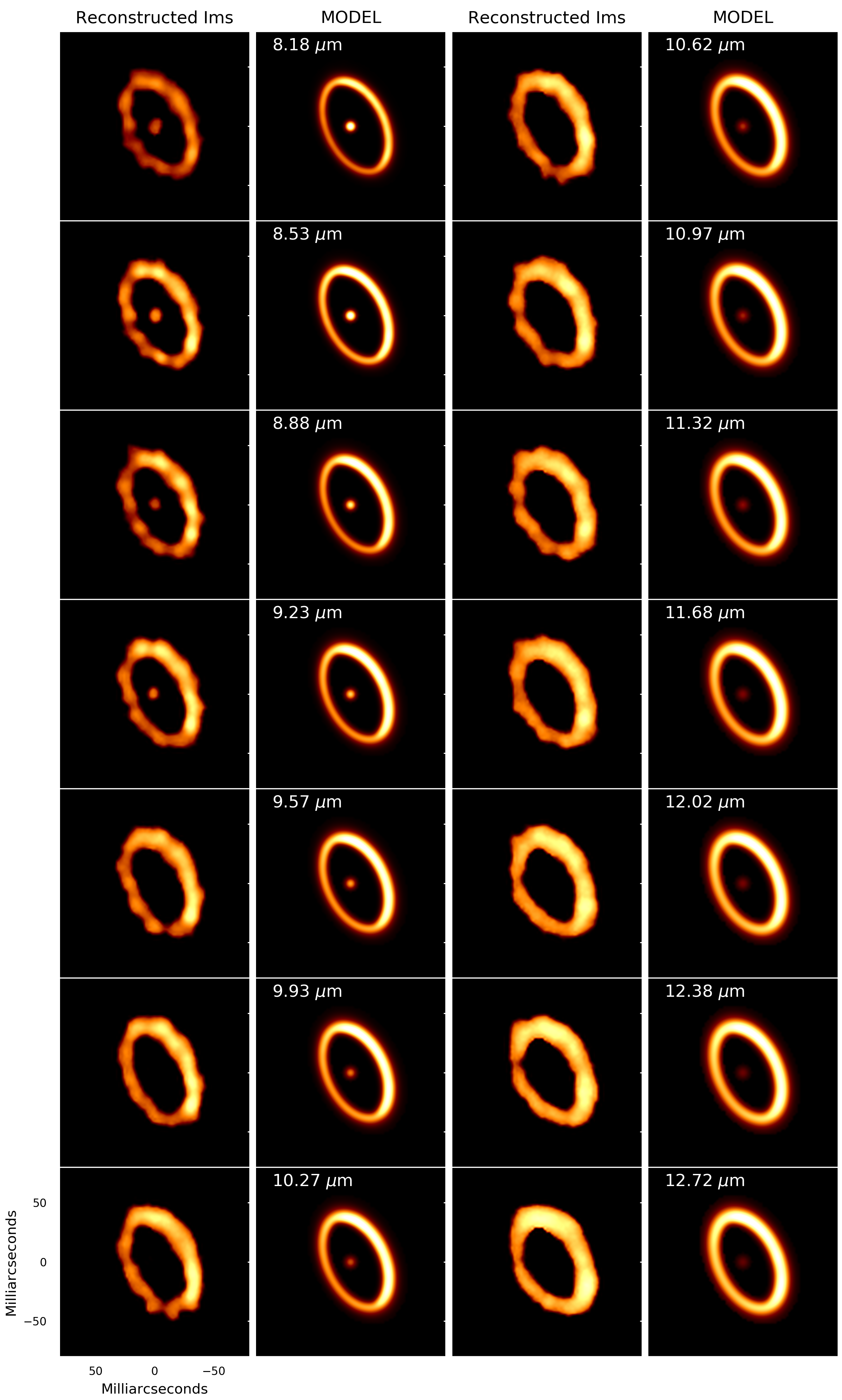}
  \caption{Best reconstructed polychromatic images. The figure shows 14 recovered images that correspond to each one of the channels sampled with the simulated data. For comparison, each one of the model images are also plotted. The wavelength of each image is shown in the model frames. The colormap of the images is normalized to the maximum pixel value in the model frames. Since the amount of flux in the central source is considerably smaller than the flux in the ring, the central object is only recovered in the first four frames, for the rest of the images, the structure of the ring is dominant. }
\label{fig:chromatic}
\end{figure}

While the monochromatic images are crucial as first step to the reconstruction, one of the main goals of the new generation of infrared interferometers is to recover
the morphological changes of the astrophysical objects along the bandpass of the observations.
This aspect is particularly important for MATISSE, which will have a bandpass as large as $\Delta \lambda \sim$ 5
$\mu$m in the N-band. Therefore, we explored this capability by performing a polychromatic reconstruction by including the differential phase information of the data. The initial setup of the
reconstructed images considers the best image from the first monochromatic reconstruction as
starting point. In the previous reconstructions, the L0-norm and the TV regularizers exhibited
the best performance. Therefore, for this new reconstruction, both of them were used together
with a transpectral L2-norm regularization. Transpectral in this case means that the regularization was applied in the wavelength direction as well; more information about the implementation can be found in the SQUEEZE documentation. The hyperparameters were selected by tuning them manually
with values close to the best ones obtained from the L-curve analysis of the monochromatic case.
Nevertheless, we are aware that selecting the optimal ones from a multi-dimensional L-curve of
the used regularizers can optimize these values. Fourteen images were recovered; each one of
them corresponding to one of the simulated spectral channels. Figure \ref{fig:chromatic} displays the recovered
chromatic images. It is clear that, with this initial setup, the rim morphology was reproduced at all the
reconstructed channels. However, the central source was only recovered in the first four of
them . It is important to mention that the total flux in the central source corresponds to only a
small percentage of the total flux in the object. Even for the first spectral channel at 8.18 $\mu$m, it
only corresponds to 5\% of the total flux, decreasing for longer wavelengths up to $\sim$0.8\% at 12.72 $\mu$m. 

\section{Conclusion}
\label{sec:conclu}

\begin{itemize}
\item The recovery of milliarcsecond resolution interferometric images in the infrared will
represent a major breakthrough for the coming generation of beam-combiners. For
example, on the one hand, MATISSE will allow us to image astrophysical objects in the mid-infrared with
unprecedented resolution, representing a tremendous advantage with respect to its
predecessor MIDI, which only allowed for parametric modelling of the interferometric
data. On the other hand, GRAVITY will enable the possibility to use referenced-phase information by using the so-called dual-field mode. Even the James Webb Space Telescope will include a Sparse Aperture Masking mode on board, which will require mature infrared image reconstruction algorithms to recover the brightness distribution of the imaged objects. 

\item Our current understanding of the image reconstruction problem and the current
developed software allowed us to perform both monochromatic and polychromatic
image reconstructions of simulated interferometric data. For the example here presented, we
could recover the different components of a proto-typical young stellar object. However,
we have also shown that image reconstruction is still not trivial and required a
systematic study of the parameters used in the reconstruction, particularly, of the
different regularizers and the value of the hyperparameters. Therefore, it is necessary to
compare the results of the image reconstruction with several software and methods to
better understand the systematics. To accomplish this task, a standarization of the benchmarking methods should be done to properly evaluate the outcome of the different methods in a systematic way. 

\item The better we understand the requirements to achieve a science-grade images from
interferometric observations, the easier it will be to provide tools and procedures to the
community to make more accessible the use of the current techniques. This is a task that
should be addressed in the coming years as part of an effort to broaden and engage the
field with more members of the international community.

\item Testing the image capabilities of the different imaging algorithms is essential to have a full description of them and it is essential for the future of infrared interferometry in Europe.
\end{itemize}

\bibliographystyle{spmpsci}      
\bibliography{Paper_lib2}

\begin{appendix}
\clearpage


\section*{Supplementary material}

# APPENDIX

## A - INTRODUCTION

This cookbook is intended to supplement the article "Why Chromatic Imaging Matters" with more details and examples on the installation and usage of specific image reconstruction softwares.

## B - IMAGE RECONSTRUCTION INTERFEROMETRIC DATA SETS

### B.1 INTRODUCTION

In this chapter different tools and data sets, useful for interferometric image reconstruction are presented. In we start by introducing the publicly available software `ASPRO2`. This software allows the generation of synthetic interferometric data sets tailored to different instruments/observatories. The images can be generated from a few built-in functions or by importing a file. The interferometric data sets generated by `ASPRO2` are very precise as they include, e.g., shadowing in the $uv$-space generations, as well as noise models of the different instruments. Then, we present the Jean-Marie Mariotti Center for Expertise in Interferometry (JMMC) and the International Astronomical Union (IAU) Interferometry Imaging Beauty Contests (§B.4). Finally, in §B.5, a library allowing the generation of synthetic images (which can then be uploaded to `ASPRO2`) is presented. This library complements the possibilities of `ASPRO2` built-in functions.

### B.2 GENERATING SYNTHETIC DATA SETS WITH `ASPRO2`

`ASPRO2` (A Software to PRepare Observations) is a program developed at the Jean-Marie Mariotti Center for Expertise in Interferometry to allow users in the community to prepare interferometric observations by simulating the response of different interferometric arrays. `ASPRO2` uses a simple Graphical User Interface (GUI) in which the user can introduce his/her observing constraints, for example:

- Users can search for their targets. The tool is connected to the SIMBAD database, so that the user can look for a desired target just by specifying a name.
- Users can define the observational setup. `ASPRO2` allows to select an observing date, the instrument and interferometric configuration. Once the configuration is selected, users are able specify the integration time and the number of position hours ($u - v$ coverage) required.

With a given setup, `ASPRO2` generates the amplitude and phase of the visibilities, the powerspectrum (squared visibilities) and the argument of the bispectrum (i.e., closure phases). In fact, the user is able to export the generated interferometric observables

into standard `OIFITS` files that can be used for modeling and image reconstruction. For a complete information of the `ASPRO2` capabilities please see: Duchene et al. (2004); Duvert et al. (2002); Mella and Duvert (2004).

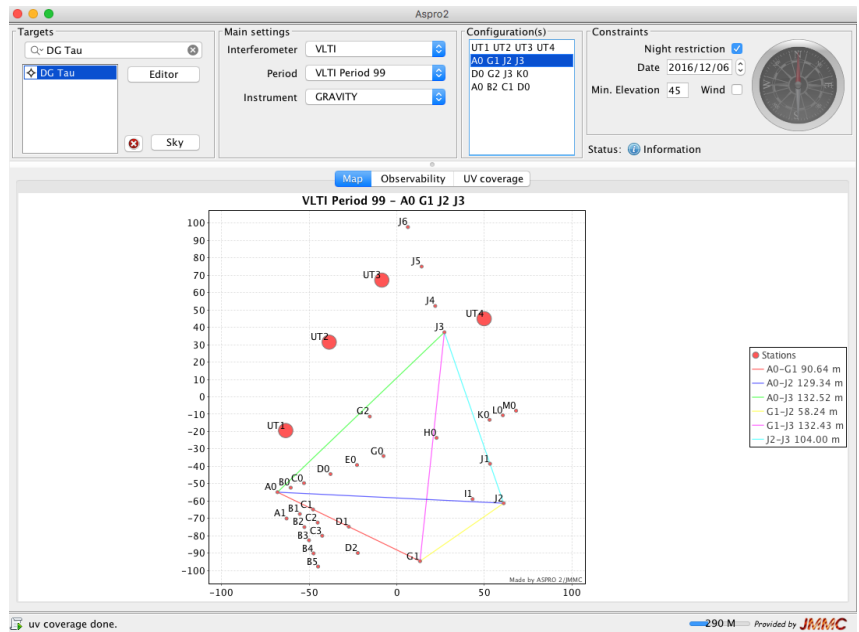

Figure 1: The figure displays the main ASPRO2 window. In the upper-left corner the user can specify the name of the target(s) to be analyzed. The upper-middle selection fields allow to define the instrumental configuration and observational ESO period. The upper-right panel serves to select the date of the observations and minimum elevation of the target. The lower section displays can show the interferometric array and baselines, the observability of the source over the night or the $u - v$ coverage.

## B.3 MODELS FOR IMAGE RECONSTRUCTION

Inside the `Editor` menu of the `ASPRO2` interface, users can corroborate and edit some of the main properties of the target, like the magnitude in different observing bands, the proper motions, coordinates, spectral type, etc. However, they can also define either a geometrical model or a user-defined model of the object's morphology. Figure 2 displays a screenshot of the target editor in `ASPRO2`. Highlighted in red, there is the sub-panel in which the user can define a model for the selected target.

### B.3.1 GEOMETRICAL MODEL

Inside the `Model` panel, the user can select over different components to simulate the target's morphology. Each one of them is composed by a given number of parameters that the user may adjust in order to cover his/her needs. For example, in Figure 3 we show the model of a source that is composed by two components: a central source (Gaussian) surrounded by an elongated ring. Both of them centered in the middle of the pixel grid. The Gaussian has a full-width-at-half-maximum of 3 mas. The ring has a minor internal diameter of 30 mas and a width of 8 mas. The elongation ratio is of 1.5 and both components have 50% of the flux in them. For more information about how to build models

with the pre-defined geometrical functions, the user can consult the ASPRO2 user manual[1] accessible thought the Help panel in the Software. Once the target model is defined, the user can check the $u - v$ coverage and the expected interferometric observables in the ASPRO2 main window. Figure 5 shows the interferometric observables obtained with the aforementioned model for UT observations with GRAVITY in low-resolution mode (see Figure 4).

### B.3.2 USER-DEFINED MODEL

As well as the geometrical model, ASPRO2 also allows the user to upload a pre-defined model. The model can be either (i) a monochromatic image or (ii) a set of images in a data cube. The images should be included in FITS files. The user can enable this option inside the Model menu, by selecting the option User Model. In order to read correctly the image(s) inside ASPRO2, the user must corroborate that the FITS header contains the keywords displayed in Figure 6. Notice that the CDELT1 should be negative it, by default, if the simulated image East was defined to the left. In the case the user uploads a data cube, CDELT3 defines the increment in wavelength for each one of the frames in the cube, CRVAL3 defines the wavelength at which the simulated bandpass begins and CUNIT3 sets the units in which CDELT3 and CRVAL3 are defined.

Once the data cube is upload, the user can check that everything is correct in the display panel inside the Model editor menu. If a data cube is uploaded, ASPRO2 animates the morphological changes of the source across the simulated band. Figure 7 displays an example of the ASPRO2 plot with the disk model used for the Beauty Contest 2016 was added. After the image was uploaded, similar as with the geometrical model, the user can adjust the desired $u - v$ coverage and the interferometric observables. Finally, if the simulation fulfills the desired requirements, the user can save it in a standard OIFITS file. To do this, the user has to select the File menu and then the Export to OIFits file(s) option.

---

[1] Also available here: http://www.jmmc.fr/twiki/bin/view/Jmmc/Software/JmmcAspro2

4

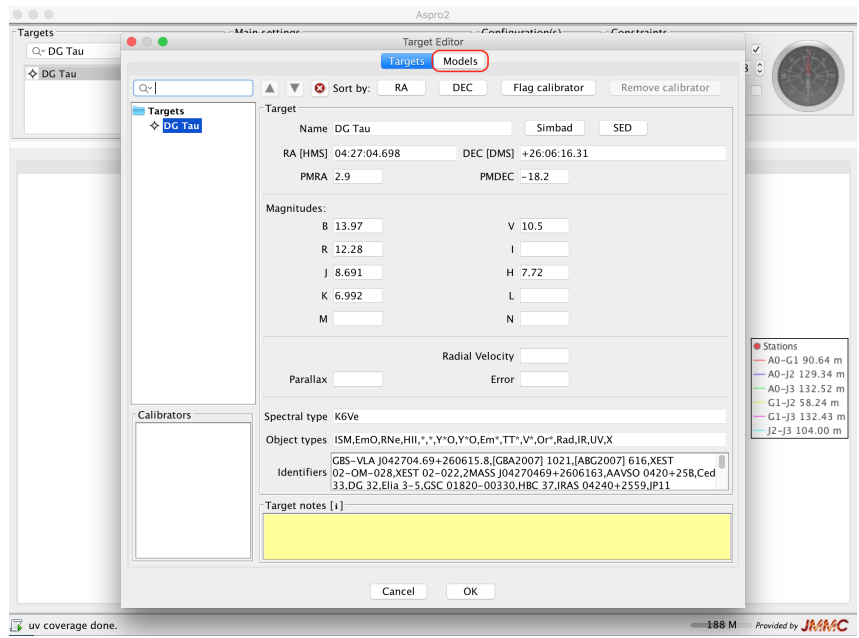

Figure 2: The figure displays the target editor menu inside ASPRO2.

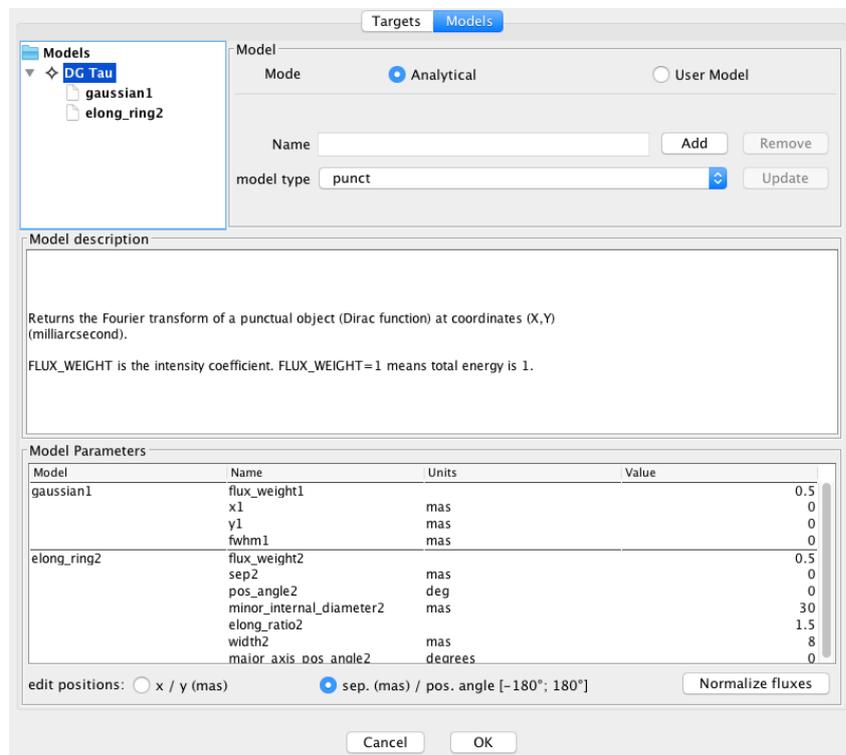

Figure 3: The figure displays the geometrical model menu of ASPRO2

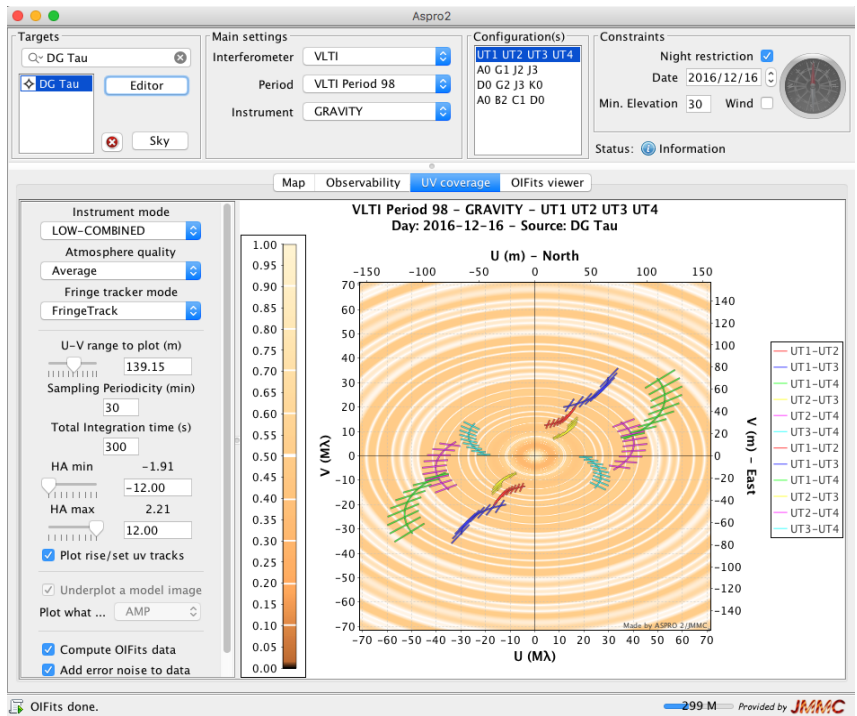

Figure 4: The figure displays the $u - v$ coverage display as well as the instrument setup panel of ASPRO2.

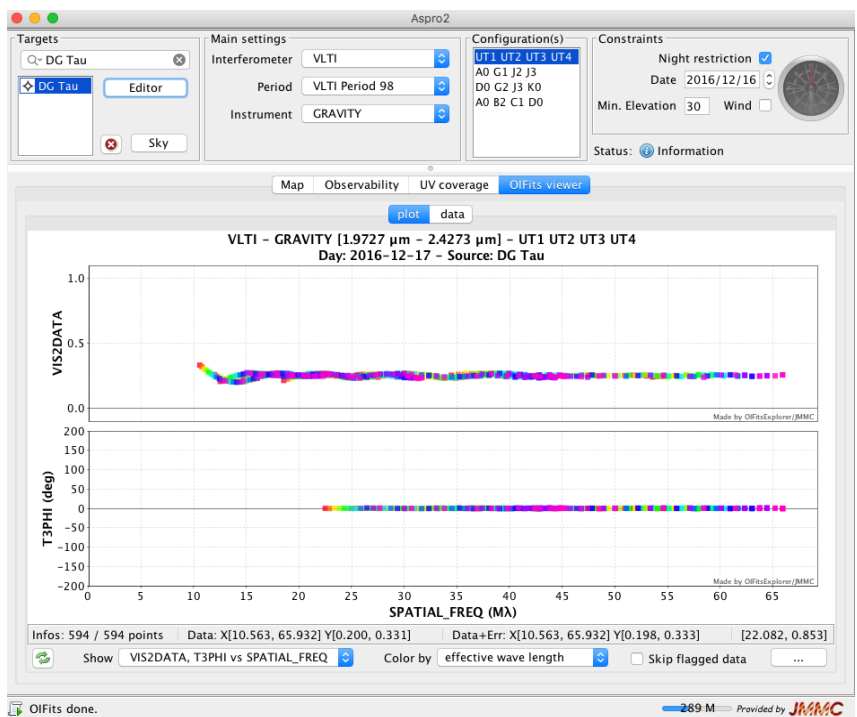

Figure 5: The figure displays the ASPRO2 plots with the simulated interferometric observables.

```
SIMPLE  =                    T / conforms to FITS standard
BITPIX  =                  -64 / array data type
NAXIS   =                    3 / number of array dimensions
NAXIS1  =                  420
NAXIS2  =                  420
NAXIS3  =                  250
CRPIX1  =                210.0
CRPIX2  =                210.0
CDELT1  =              -0.0001
CUNIT1  = 'ARCSEC  '
CDELT2  =               0.0001
CUNIT2  = 'ARCSEC  '
CRVAL1  =                  0.0
CRVAL2  =                  0.0
CRPIX3  =                    1
CDELT3  = 2.40963855421682E-09
CRVAL3  =              1.9E-06
CUNIT3  = 'METER   '
END
```

Figure 6: The figure displays the header keywords required to successfully upload a FITS data cube (with a user model) to ASPRO2.

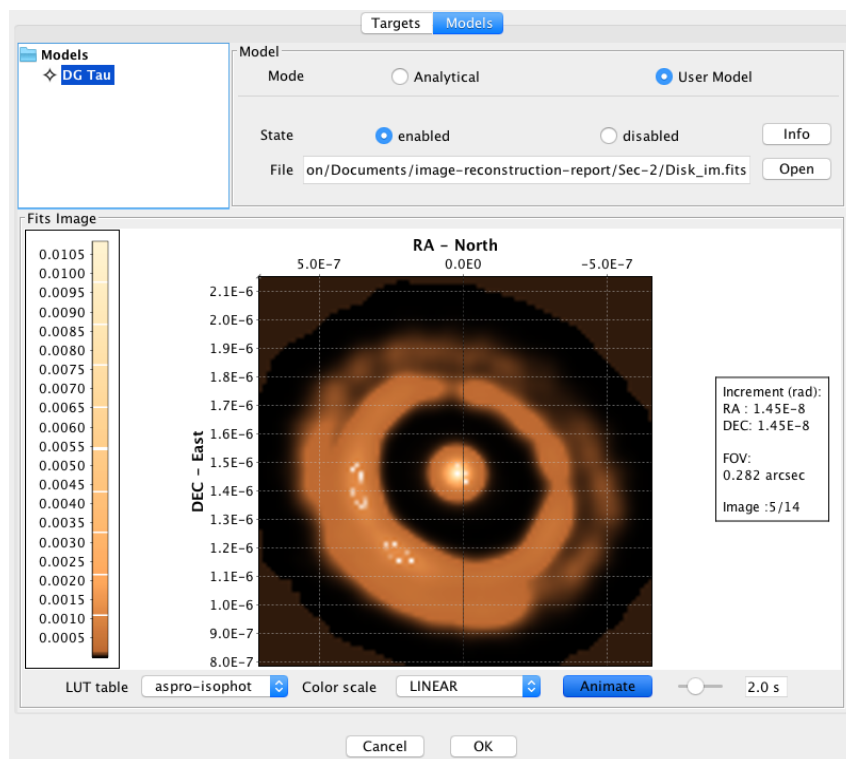

Figure 7: The figure displays the ASPRO2 panel with after a user-defined model has been uploaded using a FITS file.

## B.4 THE JMMC AND BEAUTY IMAGE LIBRARIES

The page <http://apps.jmmc.fr/oidata/> keeps several optical interferometric datasets, to be shared with the community. The main goal is to make available a repository of real or simulated reduced data, in order to practice various tools, such as image reconstruction packages as the ones described in this cookbook. Included are the data used in past IAU Interferometry Imaging Beauty Contests (from 2004 to 2010). The maintainers of the repository are continuously accepting new entries and the reader should directly visit the aforementioned link for a list of available objects and details about them.

## B.5 THE YORICK SYNTHETIC IMAGE LIBRARY

In addition to the `Aspro2` built-in functions to generate astronomical source brightness distributions, there is a repository available at GitHub, named YSIL (Yorick Synthetic Image Library), with `Yorick` code that allows the creation of basic objects. This library can be used to generate astronomical synthetic sources, such as stellar clusters, young stellar objects and stellar photospheres (e.g., see Gomes et al., 2017, for some examples). The code can be found at <https://github.com/NunonuN/ysil>. It is open source under the GNU General Public License version 3, and it is in its final pre-release state.

The following subsections briefly explain how to install and use YSIL. For further details, please visit the aforementioned website.

### B.5.1 INSTALLATION

In order to use YSIL, it is necessary to install `Yorick` (see F.2, page 39 for details). Download YSIL to any desirable folder and include `ysil.i` in `Yorick`, as in, for example

```
include, "ysil.i";
```

### B.5.2 EXAMPLES

Some examples of simple structures or objects created with YSIL are illustrated in fig. 8. The uniform discs were created with the following command:

```
x= span(-10, 10, 300)(, -:1:300);
discs= ysil_uniform_disc(x,
                    [[2, 4], [3.5, 3.5], [5.6, 1.3]], // radi
                    [[-6, 0], [5, 5.5], [4, -7]], // centrs
                    angs= [30, 0, -160]);
```

the Gaussian disc was obtained with

```
x= span(-10, 10, 500)(, -:1:500);
disc= ysil_gaussian_disc(x, [4., 2.], angs= 10);
```

the Gaussian rings were built with

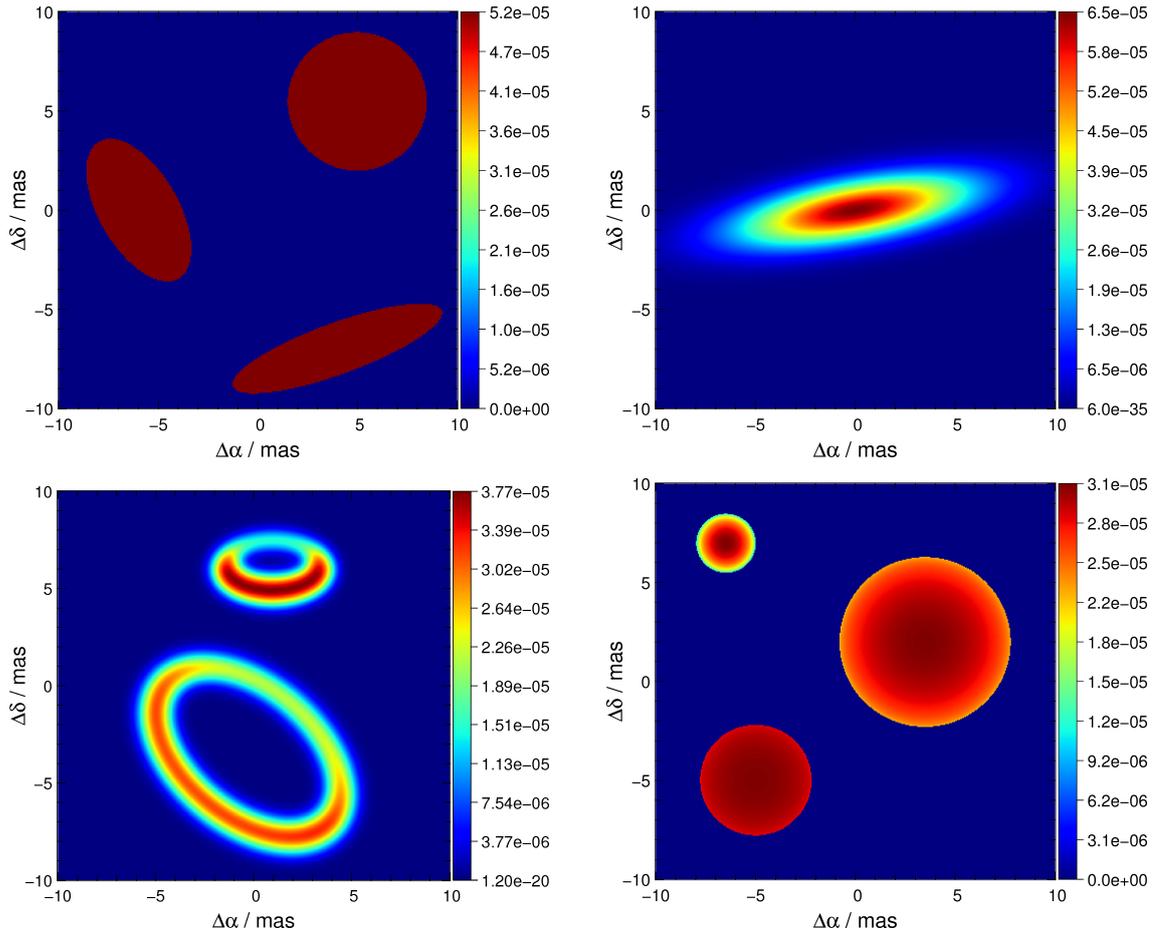

Figure 8: *From left to right, top to bottom*: three uniform discs, a disc with Gaussian profile, two rings with Gaussian profile, and three stellar photospheres, all created with the YSIL package, in a FOV of $20\,\mathrm{mas}^2$.

```
x= span(-10, 10, 500)(, -:1:500);
rings= ysil_gaussian_ring(x,
                    [[6.1, 3.8], [3.0, 1.7]], // out_radi
                    [[5.0, 2.9], [2.0, 0.9]], // in_radi
                    [[-0.3, -3.5], [1.0, 6.0] ], // centrs_out
                    [[-0.2, -3.3], [1.0, 6.45]], // centrs_in
                    [0.5, 0.4], // sigs
                    angs= [-40.0, 0.0]);
```

and the stellar photospheres with

```
x= span(-10, 10, 400)(, -:1:400);
stars= ysil_limb_darkened_disc(x,
                    [2.8, 4.3, 1.5], // radi
                    [[-5, -5], [3.5, 2], [-6.5, 7]], // centrs
                    u= [0.1, 0.3, 0.7]);
```

# C - IMAGE RECONSTRUCTION WITH SQUEEZE

## C.1 INTRODUCTION

SQUEEZE is an open source (GPL v3) image reconstruction software developed by Fabien Baron[2] at Georgia State University. It uses a Monte-Carlo Markov Chain (MCMC) approach as main minimization engine. Two type of MCMC algorithms are included: Simulated Annealing and Parallel Tempering, both of them with Metropolis-Hasting moves. SQUEEZE includes a full polychromatic imaging, performing a simultaneous model-fitting to: (i) visibilities (amplitudes and phases), (ii) powerspectra, and (iii) bispectra, or combination of them. The code has the flexibility to consider the visibility phases as differential ones. It works either with `OIFITS V1.0` or `OIFITS V2.0` tables. The official description of the algorithm can be found in Baron et al. (2010), while the source code and its documentation are reachable at the following link: https://github.com/fabienbaron/squeeze.

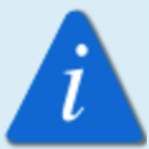
**Disclaimer**

The current cookbook has been written from a user perspective. The main aim of the current document is to provide an guide of the main characteristics of SQUEEZE following practical examples. This document does not attempt to replace the official description and software documentation. If SQUEEZE is used or modified, the corresponding original papers and documentation should be always cited according to the established copyright.

---

[2] e-mail: baron@chara.gsu.edu

## C.2 INSTALLATION AND CONFIGURATION

SQUEEZE is a stand-alone package that can be retrieved from the following GitHub repository: https://github.com/fabienbaron/squeeze. The code is written in C and it is compatible with OpenMP. The code requires `gcc` to be compiled. This compiler is native in most of the Linux distributions. Nevertheless, it has to be installed by the user in OSX. The GitHub repository contains all the necessary instructions to install these compiler. However, the most reliable ways to do it on an OSX system is through the `macports` [3] or the `Homebrew`[4] environments. According to the source repository, once the compiler and dependencies are satisfied, the SQUEEZE installation is relatively simple, a sequence of the following commands has to be executed:

- Download the current version of the code from the host repository or use the following `git` command:

```
git clone https://github.com/fabienbaron/squeeze.git
```

- Go inside the SQUEEZE main directory and update the OIFITSLIB module with the following command:

```
git submodule update --init
```

- Inside the main SQUEEZE directory use the following `cmake` commands to build the executable:

```
cd squeeze/build
cmake ..
make
```

- Once the executable is created the user can add it to the global path. For a bash shell use: `export PATH=/path/to/SQUEEZE main dir/bin $PATH` For a tsch shell use: `setenv PATH $PATH:/path/to/SQUEEZE main dir/bin`

If everything went well, the user should be able to execute squeeze from a terminal session. A simple test to identify any problem in the installation is to type the following command in the prompt:

```
squeeze -h
```

This command should list all the SQUEEZE options available to the user. These options are divided into six main categories: (i) Main image settings, (ii) Output settings, (iii) Simultaneous model fitting settings, (iv) Oifits import settings, (v) Regularization & init settings and (vi) Convergence settings.

---

[3] https://www.macports.org/
[4] http://brew.sh/index_es.html

## C.3 BASIC USAGE

One of the main advantages of SQUEEZE versus other reconstruction packages is the large number of options and parameters that can be used. These characteristics offer a large flexibility for the reconstruction, but it also requires more training from the user to master all the different options. In this cookbook, some examples of the basic code usage are included, as well as more advanced ones. **Nevertheless, the user has to be aware that image reconstruction is not a straight forward process and may require a deep understanding of the different parameters involved.**

### C.3.1 EXAMPLE 1

For the first example, the model of LkHa 101 used for the 2004 Beauty Contest (Lawson et al., 2004) has been selected. The data set was built assuming observations with the six-station Navy Prototype Optical Interferometer (NPOI). Figure 9 displays the true image used to simulate the data set as well as the simulated $u-v$ coverage. The data set consisted of 15 baselines tracked at 15 different hour-angles. For this example, the OIFITS file only consisted of squared visibilities and closure phases.

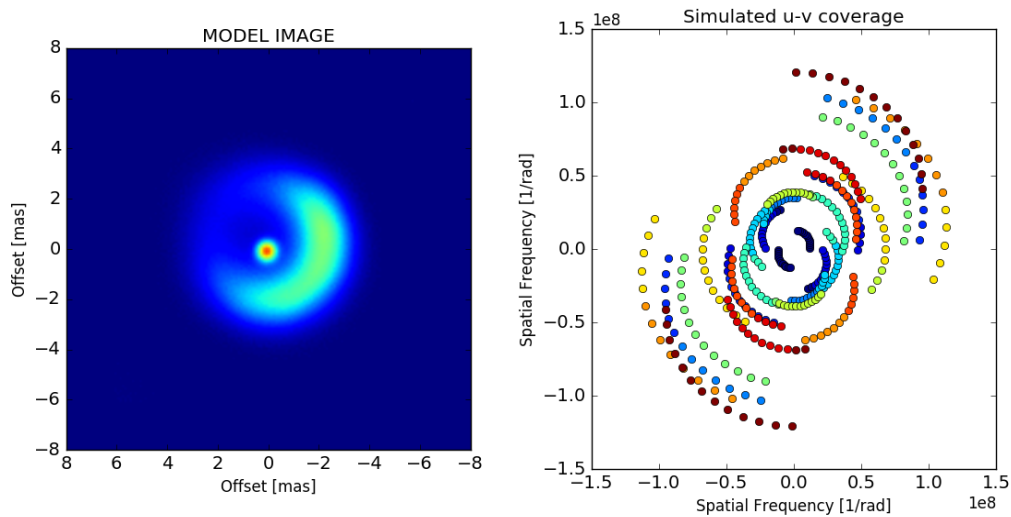

Figure 9: The **left:** panel displays the model image of LkHa 101 used to simulate the NPOI visibilities. The **right** panel displays the $u-v$ coverage of the interferometric data. Each different color indicates a different baseline.

#### C.3.1.1 RECONSTRUCTION WITHOUT REGULARIZATION:

To begin with a reconstruction, the pixel grid and pixel size should be defined by the user. As a consequence of mapping the interferometric data into a discrete grid of pixels, the user should be aware of avoiding field-of-view aliasing, at the time the field-of-view is chosen large enough to sample all the measured spatial frequencies in the data. As a rule of thumb, the pixel size of the reconstructed image should be of the order of:

$$\Delta\theta \leq \frac{\lambda}{4\text{B}_{\text{max}}} \qquad (1)$$

where $\text{B}_{\text{max}}$ corresponds to the maximum baseline length in the interferometric array. In this case, the wavelength, $\lambda$ corresponds to $0.55\mu$m and the maximum baseline length corresponds to 66.4 meters. Therefore, the pixel size of the reconstructed image should be of the order of $\Delta\theta \sim 0.4$ mas. The following SQUEEZE command allows the reconstruction of an image of 128×128 pixels with the derived pixel scale using the data set called `example1.oifits`. In this case, no regularization functions were selected and only one MCMC chain with 500 iterations (default option) is used:

```
squeeze example1.oifits -w 128 -s 0.4
```

Figure 10 displays the reconstructed output images from SQUEEZE. Three panels are observed, which corresponds to the mean, mode and median images obtained from the iterations after the MCMC burn-in phase. In this case, the mean image was computed only with the best 35% of the frames (i.e., with the lowest $\chi^2$) in the MCMC chain. If the algorithm does not converge (i.e., it does not pass the burn-in phase in a given number of iterations), SQUEEZE still returns the aforementioned three images but they might have a large number of artifacts, particularly "image ghosts" due to the misaligned frames at different chain iterations.

The SQUEEZE output images are standard `.fits` files and the user could use his/her prefered software to inspect them. Additionally to these files, SQUEEZE can deliver the changes in the posterior probability at each one of the iterations in the chain. To habilitate this option, the user could run the following command as an example:

```
squeeze example1.oifits -w 128 -s 0.4 -fullchain
```

The results of the probability changes are thus saved in a file called `output.fullprobs`. This file contains the following information:

- The number of chains and the number of iterations
- The objective temperature[5]
- The logarithmic likelihood ($\chi^2$), the prior probability and the posterior probability for each one of the iterations in the different chains.

Figure 11 displays the changes of the likelihood (in log space), prior probability and posterior probability for the previous example. Notice that, although the reconstruction did not use a prior function, the code produces a prior probability because by default SQUEEZE uses the **field of view** regularizer to move the image to its centroid position at each one of the iterations in the MCMC. In the previous example, it is clear that the effect

---

[5]If the user runs several simultaneous chains, the total number of chains should be divided by three and this will be the number of lines that will contain the objective temperature before the posterior probability information.


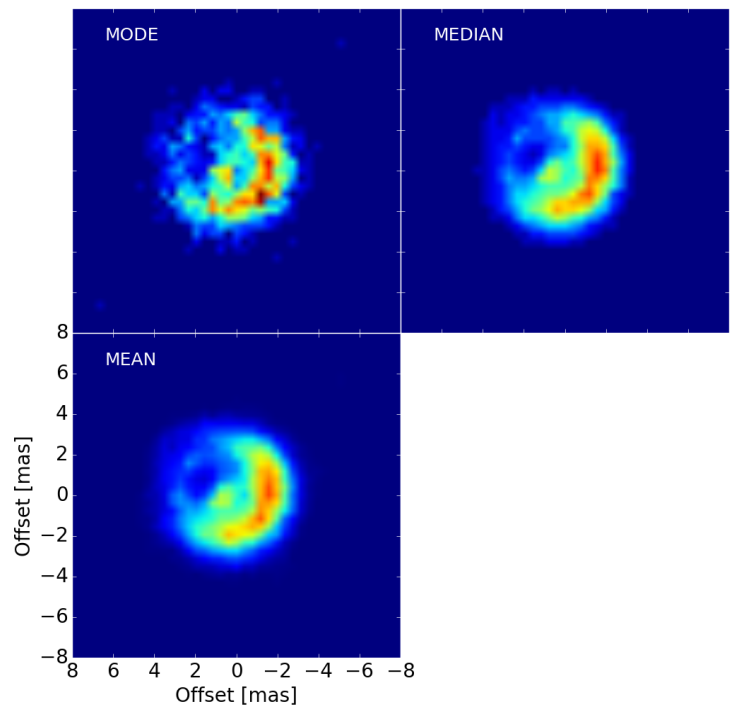

Figure 10: The figure displays three panels that correspond to the reconstructed images delivered by SQUEEZE after a given number of iterations for one MCMC chain. The color scheme is the same in all the frames, therefore the user can identify the differences between them.

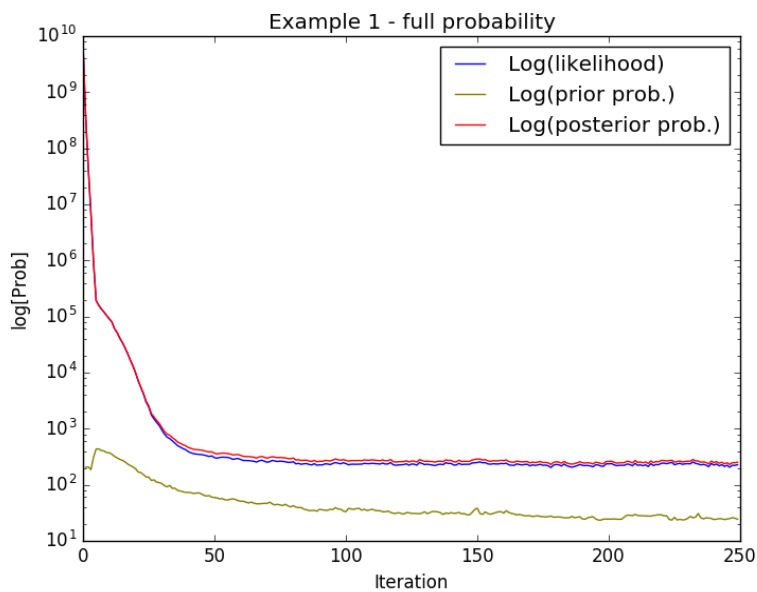

Figure 11: The figure displays a plot of the probability (in log space) changes for each one of the iterations in the MCMC. The likelihood, prior probability and posterior probability are displayed with different colors (see label in the image). Notice how after iteration number 50, the value of the posterior probability is quasi-constant around ~250, which corresponds to a total reduced $\chi^2_r$ ~1.04, indicating that the code has properly converged.

of the prior probability is almost negligible once the algorithm has converged. Therefore, the posterior probability is mainly affected by the likelihood (i.e., the difference between the interferometric observables and the model). Visualizing the `.fullprobs` file is particularly interesting to identify whether the algorithm is properly converging or not. Additionally to the `.fullprobs` file, SQUEEZE also generates a file that contains the reconstructed image at each iterations of the MCMC and it is specially useful to make a statistical analysis of the frames selected for the final image. This file is usually named `output_fullchain.fits.gz`.

### C.3.1.2 RECONSTRUCTION WITH REGULARIZATION:

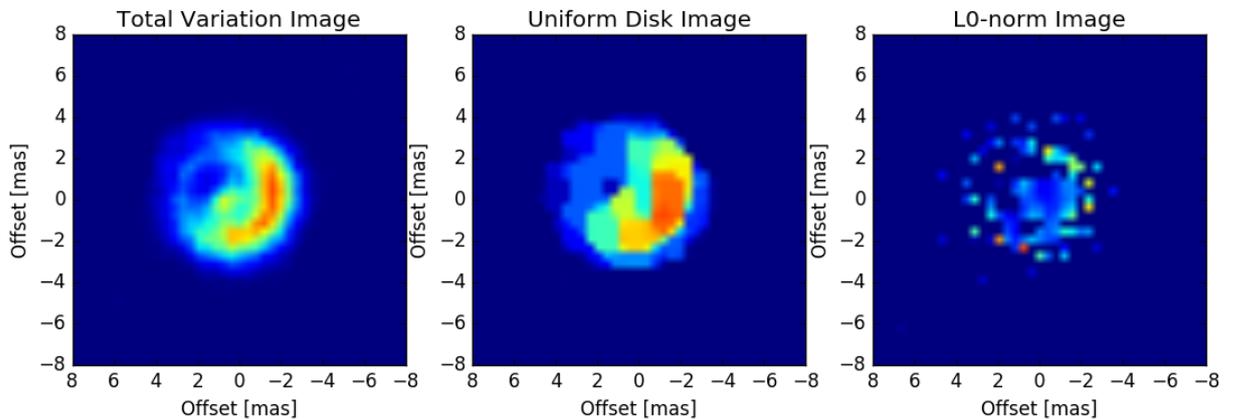

Figure 12: The image displays three panels that correspond to the reconstructed images with three different regularizers (see labels in the image). The color scheme is the same among them.

Now that the basic reconstruction with SQUEEZE has been described, it is time to test the reconstruction using some regularization functions. Selecting a regularizer is not an easy task and strongly depends on the "prior" knowledge that the user has of the source's brightness distribution. There would be targets for which the structure is well-known (e.g., a binary) or that could be inferred from a parametric model (e.g., a disk or gaussian envelope), but there would be others more difficult to deduce. SQUEEZE allows to use several regularization functions simultaneously. Nevertheless, the more regularizers are used, the more difficult is to estimate the optimal values of the hyperparameters. To reconstruct the image with a given regularization, the user should check the corresponding option in SQUEEZE. For example, the following commands will serve to reconstruct the image with the associated regularizers:

- Reconstruction with **total variation** and an hyperparameter value, $\mu_{\mathrm{TV}}$, of 100:

```
squeeze example1.oifits -w 128 -s 0.4 -tv 10
```

- Reconstruction with **uniform disk** and an hyperparameter value, $\mu_{\mathrm{UD}}$, of 10:

```
squeeze example1.oifits -w 128 -s 0.4 -ud 10
```

- Reconstruction with **L0-norm** and an hyperparameter value, $\mu_{L0}$, of 10:

```
squeeze example1.oifits -w 128 -s 0.4 -l0 10
```

Figure 12 displays the reconstructed images with the previous parameters and Figure 13 displays the posterior probability for each one of the iterations. Notice how, although the probability converged for the three images, the recovered image distribution is quite different among the reconstructions. **Therefore, caution should be taken by the user, when a regularization is selected.**

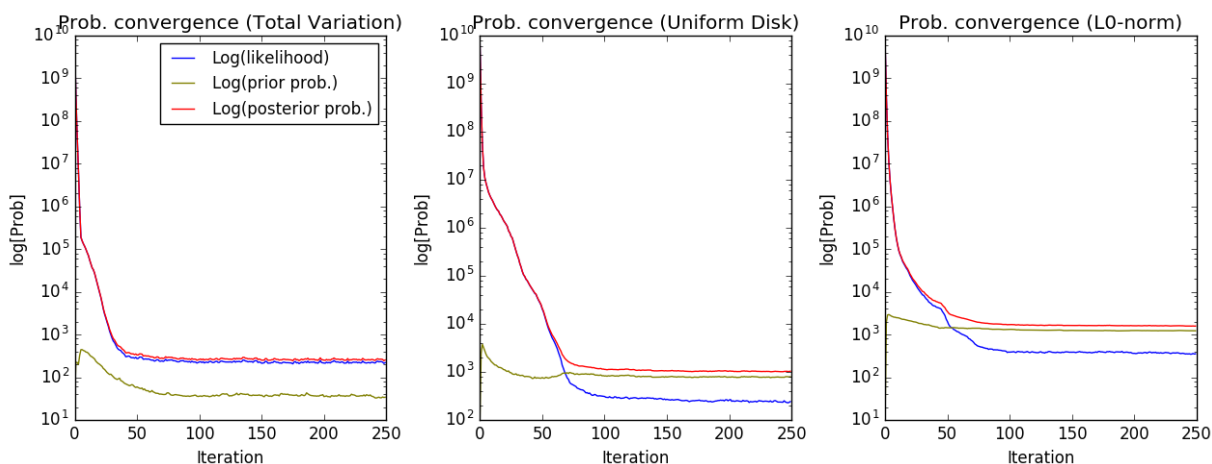

Figure 13: The image displays three panels that correspond to the posterior probability convergence with three different regularizers (see labels in the image).

### C.3.1.3 SELECTION OF THE HYPERPARAMETERS $\mu_i$

The hyperparameters $\mu_i$ control the trade-off between the $\chi^2$ and the prior information of the brightness distribution encoded in the regularizers $R_i(x)$. Therefore, selecting the appropriate values of $\mu_i$ is crucial for the image reconstruction process. One of the most common methods to select the optimum value of a given $\mu$ is the L-curve. It computes the image solution for several values of $\mu$, characterizing the response of the prior term versus the $\chi^2$. This relation shows an "L-shaped" curve (Bose et al., 2001; Hansen, 1992). For regions along the curve where the data fidelity term changes rapidly compared with the prior, the reconstructed image is in the over-regularized region. However, if the opposite is observed, the image is in the under-regularized region. The optimum value of $\mu$ is thus associated to the elbow of the L-curve.

Therefore, to select the optimum regularization, the user must have to conduct a search over several values of the hyperparameter. Figure 14 displays, as an example, the reconstructed image using the uniform disk regularization with different values of $\mu$.

Notice the change in the object morphology from the under-regularized ($\mu < 5$) to the over-regularized region ($\mu > 5$). Sometimes the L-curve exhibits more than one elbow. In cases like this, a rule of thumb is to use the optimal value that corresponds to the inflection point with the smallest $\chi^2$. Figure 15 displays the L-curve for the reconstructions presented previously.

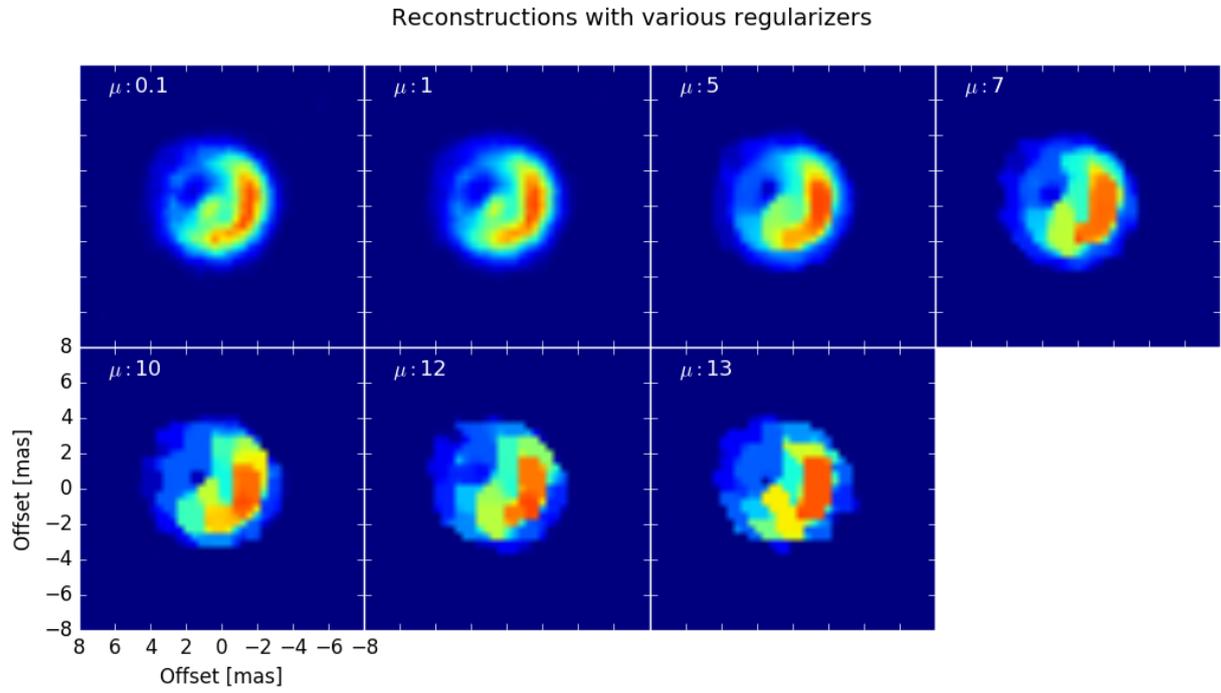

Figure 14: The image shows the impact of $\mu$ on the image reconstruction process. Seven reconstructed images of LkHa 101 are shown using different hyperparameter values. Notice the changes in the reconstructed image depending on the $\mu$ value.

C.3.1.4 SQUEEZE WITH SEVERAL CHAINS

One of the advantages of the MCMC implementation in SQUEEZE, is the possibility to perform the reconstructions using several simultaneous chains. This option is particularly important to test the outcome of the reconstruction, even if the code has converged. Analyzing the outcomes of several chains help to identify artifacts in the image and to make statistics about the most probable brightness distribution of the target. The `-chains` option uses a parallel implementation to compute simultaneously the reconstructions. The user should check what is the maximum number of chains that can be set according to the available computer facilities. As an example to habilitate this option with 50 simultaneous chains, the user could run the following command:

```
squeeze example1.oifits -w 128 -s 0.4 -tv 10 -chains 50
```

Figure 16 displays the distributions of the likelihood and prior probability for the last 100 iterations of 50 chains. Notice how both correspond to a normal distribution. The

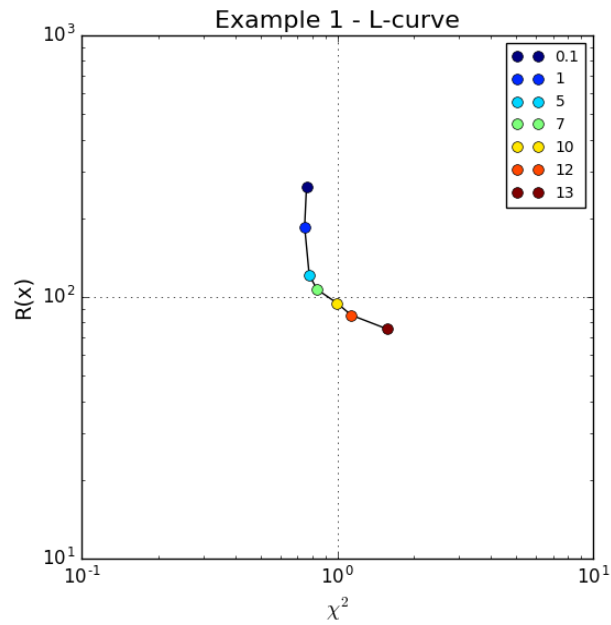

Figure 15: L-curve obtained with different values of $\mu$ using the Uniform Disk regularizer. The vertical axis displays the value of R(x) and the horizontal axis the $\chi^2$. The optimal value of $\mu$ is at the elbow of the curve with the smallest $\chi^2$, in this case it is close to 5.0.

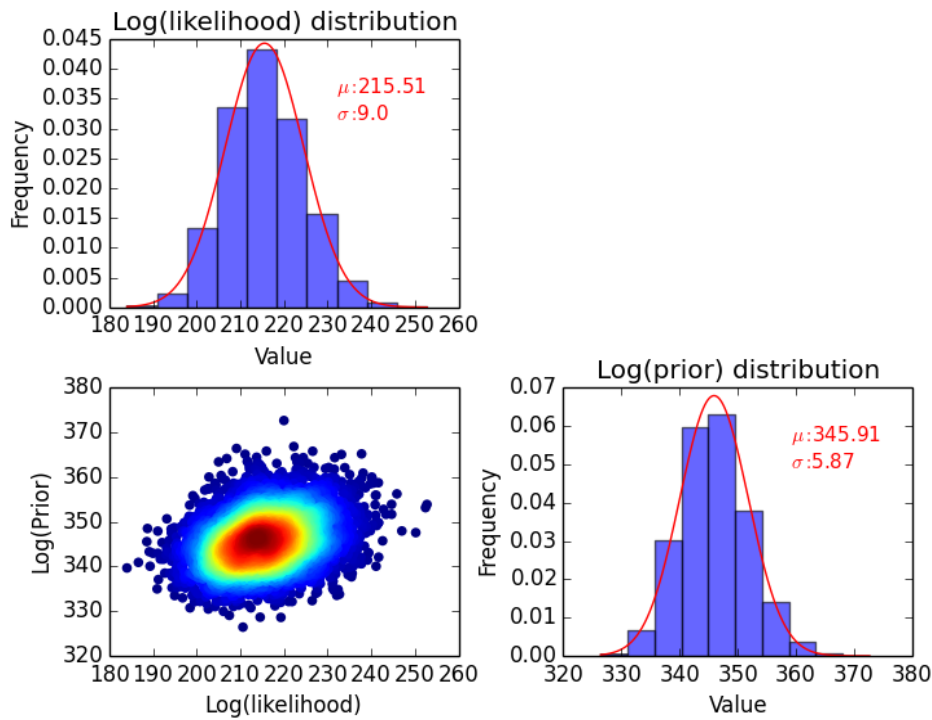

Figure 16: The Figure shows individual likelihood and prior distributions, as well as the joined distribution of 50 chains with 100 iterations.

joint distribution shows also how the different chains converged to a given value. Therefore, in order to determine what is the "best image" (the most probable one), the user has to analyze the distributions and chain outputs to decide which of them pick to create the final image. For example, in Figure 17 the mean "final image" created with the best 10 chains (i.e., with the smallest $\chi^2$) is presented together with an example of the image fit to the observed closure phases and squared visibilities.

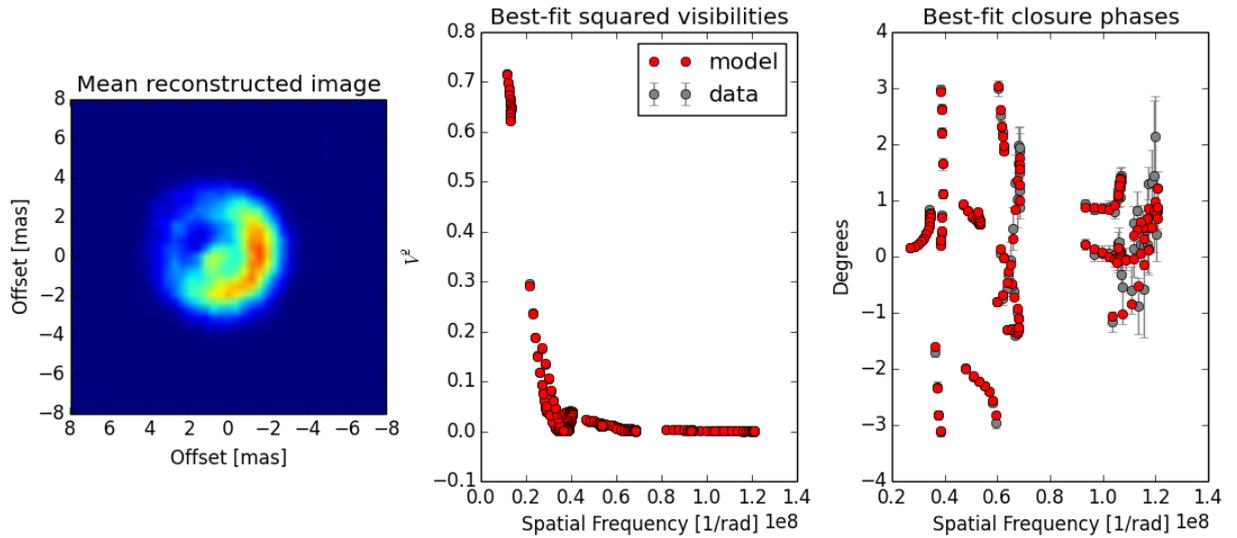

Figure 17: **left:** Best reconstructed image. **center:** Best-fit image squared visibilities. **right:** Best-fit image closure phases.

Up to now, the reconstructions have been made using using the same initial brightness distribution for all the chains. By default, SQUEEZE uses a delta function centered at the middle of the pixel grid. Nevertheless, it would be particularly important to test the reconstruction with different initial brightness distribution, specially for different chains. To do this, the user could habilitate the following options:

- Initialize the reconstruction with a random image common to all the chains:

```
squeeze example1.oifits -w 128 -s 0.4 -tv 10 -chains 50 -i random
```

- Initialize the reconstruction with a random image different for all the chains:

```
squeeze example1.oifits -w 128 -s 0.4 -tv 10 -chains 50 -i randomthr
```

- Initialize the reconstruction with a user-defined image `initial_im.fits`:

```
squeeze example1.oifits -w 128 -s 0.4 -tv 10 -chains 50 -i initial_im.fits
```

## C.3.2 EXAMPLE 2

For this example, the second data set provided for the 2004 Beauty Contest edition is used. Figure 18 displays the $u - v$ coverage, as well as the squared visibilities and closure phases for this data set. According to Lawson et al. (2004), this data set consisted of a central diffuse source together with a secondary compact source. The data used in this example not only contains closure phases and squared visibilities, but also visibility amplitudes, phases and triple amplitudes. In this respect, SQUEEZE allows the user to select which observables fit during the reconstruction. To disable observables from the minimization process, the following options could be selected: `-novis, -novisamp, -novisphi, -not3, -not3amp, -not3phi` **and** `-nov2`. The user should check the quality of the observables to decide whether use all of them or not. For example, in this case, performing the reconstruction only with closure phases and squared visibilities makes difficult the convergence of SQUEEZE, and the minimum temperature reached is of ~7.0 with a $\chi^2$ in the squared visibilities of ~3700. The user can corroborate this, by running the following command:

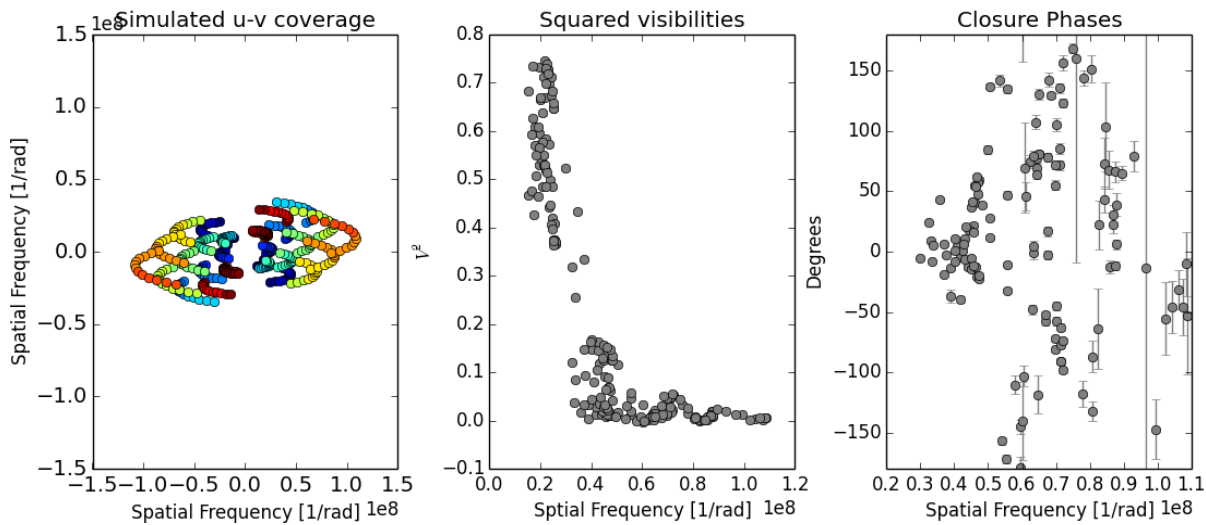

Figure 18: **left:** $u - v$ coverage. The colors indicate different baselines. **center:** Squared visibilities. **right:** Closure phases

```
squeeze 2004dataset2.oifits -w 128 -s 0.4 -tv 100 -novisamp -novisphi -not3amp
```

In contrast with the previous attempt, using the visibility amplitude and closure phases the algorithm converges reaching an annealing temperature of 1.0 and a $\chi^2$ of ~3.6. The user can check this reconstruction by running the following command:

```
squeeze 2004dataset2.oifits -w 128 -s 0.4 -tv 100 -nov2 -novisphi -not3amp
```

Figure 19 displays the reconstructed images with the previous two commands. Please notice the clear difference in the recovered brightness distribution. The left panel displays the reconstruction that converged properly and the structure of the source (an extended elongated shape + a compact object) is revealed. The right panel displays the non-converged reconstruction.

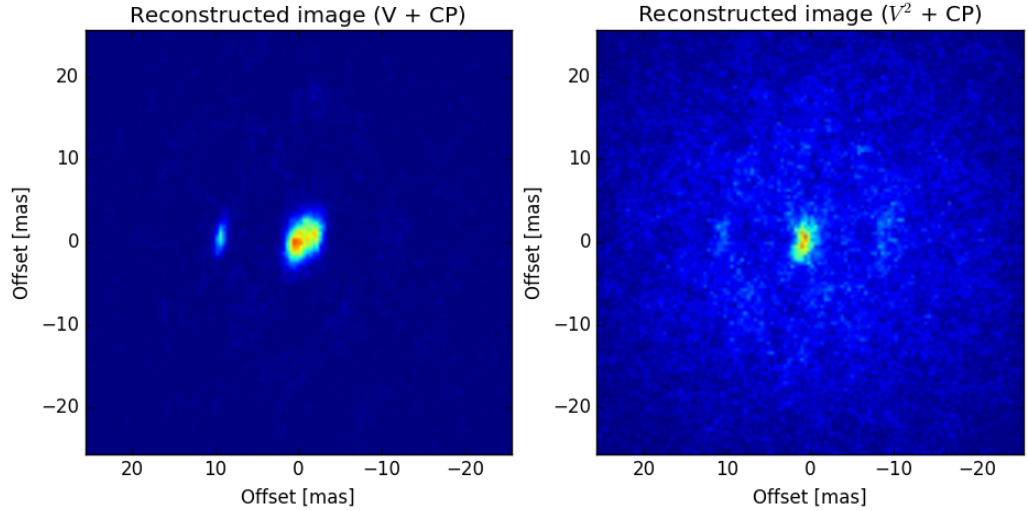

Figure 19: **left:** Reconstructed image with visibility amplitudes and closure phases. **right:** Reconstructed image with squared visibilities and closure phases

## C.4 ADVANCED USAGE

### C.4.1 POLYCHROMATIC RECONSTRUCTION

One of the main goals of the new generation of infrared interferometers is to recover the morphological changes of the astrophysical objects across the bandpass of the observations. For example, this aspect is particularly important for MATISSE (Lopez et al., 2008, 2009), which will have a bandpass as large as $\Delta\lambda$ ~5 $\mu$m in the N-band. SQUEEZE allows to perform polychromatic reconstruction by including the differential phase information of the data. The initial setup required to perform this reconstruction is quite similar to the ones used in the previous examples. Nevertheless, this time, the differential phases are included. SQUEEZE has an integrated a transpectral regularizer available through the $-ts$ option. Since closure phase is a shift-invariant observable, transpectral regularization is required to ensure spectral continuity across the bandpass used for the reconstructions. This is particularly important to avoid different positions of the recovered object at each one of the frames in the reconstructed data cube.

### C.4.2 THE OBJECT

Proto-planetary disks is one of the key science objectives for the second generation of VLTI instruments. Therefore, the object selected to illustrate a polychromatic recon-

struction consisted of a simulated accretion disk as observed in the N-band (8-13 $\mu$m) with the MATISSE instrument in low-resolution mode. The model includes a central star (surrounded by a large amount of dust) and a disk containing one gap, supposed to be carved by a forming planet (Gonzalez et al., 2015). The brightness distribution across the disk is not uniform with a clear bright spot at a position angle of ~220° East of North and the disk shape is elongated in the East-West direction. The observational setup for this data set includes three simulated nights using three different configurations of the Auxiliary Telescopes. Figure 20 displays the simulated $u - v$ coverage, as well as the interferometric observables. Notice how the squared visibilities trace a resolved object with an angular size of ~100 mas. The simulated closure phases are considerably noisier, which increases the difficulty of the reconstruction. The differential phases use the AMBER convention in which the reference channel corresponds to an average of all the channels except the working one. This means that the reference channel varies across the bandpass. The data set used was part of the 2016 Beauty Contest (Sanchez-Bermudez et al., 2016) and can be found here: www.opticalinterferometry.com

### C.4.3 THE RECONSTRUCTION

To begin with this reconstruction, an initial brightness distribution of a Gaussian was used. To include a pre-defined image, in SQUEEZE the user can use the `-i` option. In this case, the data channel with the longest wavelength (from 12.8076 to 13.20 $\mu$m) and higher signal to noise was used. For this purpose the `-wavchan` was enabled. With this setup, a first monochromatic reconstruction was performed with the following command:

```
squeeze 2016dataset2.oifits -w 135 -s 3.0 -la 1000 -l0 1 -i initial.fits -novisamp -
    not3amp -novisphi -v2s 2.0 -wavchan 0 1.26e-5 13.10e-6 -n 150 -chains 3
```

In this example, a Laplacian regularization $\mu_{Lap}$=1000 was used. Please notice that the aforementioned value was obtained by using the L-curve described in Sec. C.3.1.3. In the command, we have also disabled the use of the visibility amplitudes (not included in the data set), the triple amplitudes (also not included in the data) and the visibility phases. To ensure convergence, the additional option `-v2s` was included. This option multiply the errorbar of the squared visibilities by a factor of two. It is important to highlight, that the user should decide whether this option is necessary for a given reconstructed data set because it relaxes the convergence criteria of the algorithm. Finally, three chains were created for the reconstruction. The Figure 21 displays the initial image and the result of the monochromatic reconstruction.

Once the monochromatic reconstruction was finished. The recovered image was used as starting point to perform a full polychromatic reconstruction. The two regularizers used in the monochromatic reconstruction were included with an additional transpectral regularizer. To enable the reconstruction using all the channels in the data

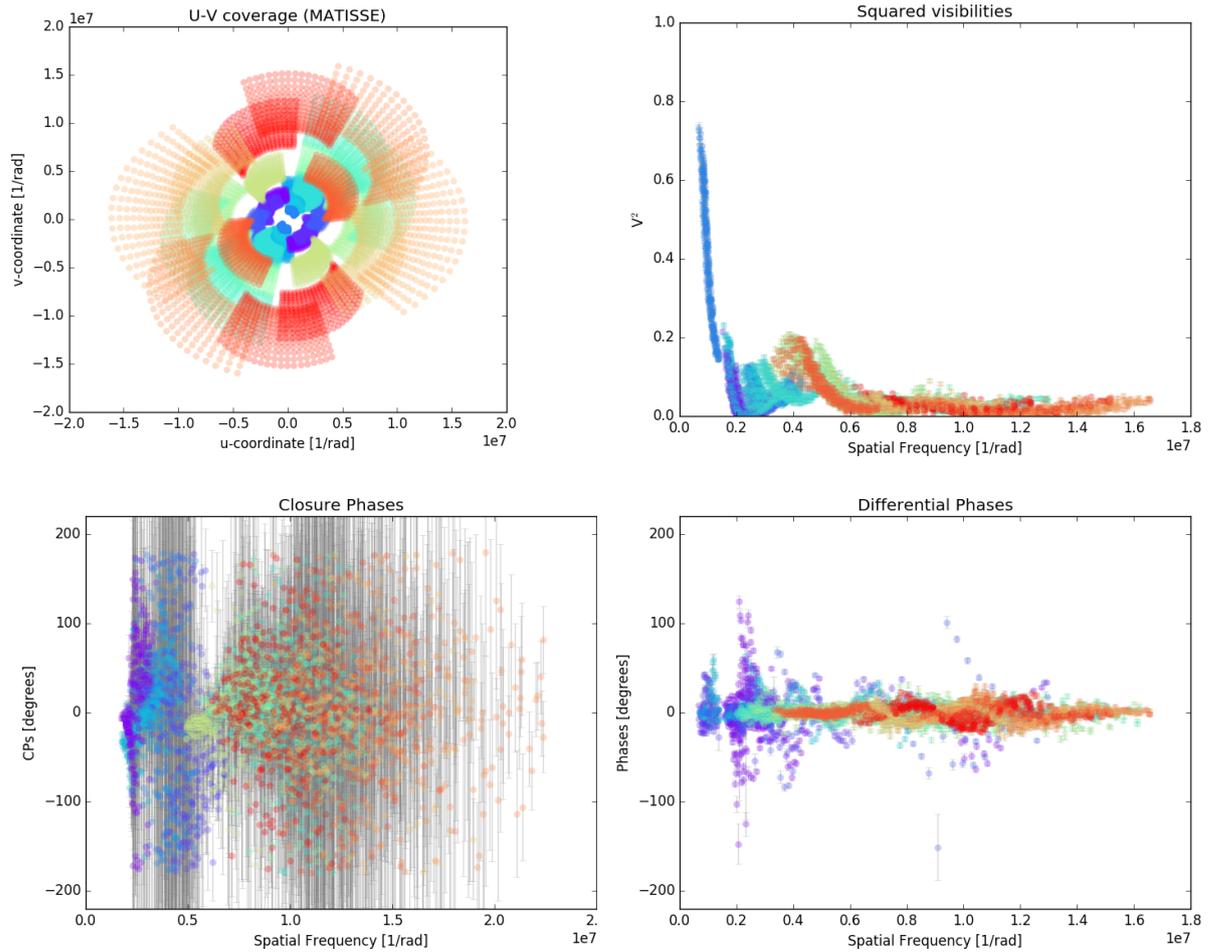

Figure 20: The **upper-left** panel shows to the simulated $u-v$ coverage for the MATISSE data. The **upper-right**, **lower-left** and **lower-right** panels display the simulated closure phases, squared visibilities and differential phases respectively. Different colors correspond to different baselines or triangles.

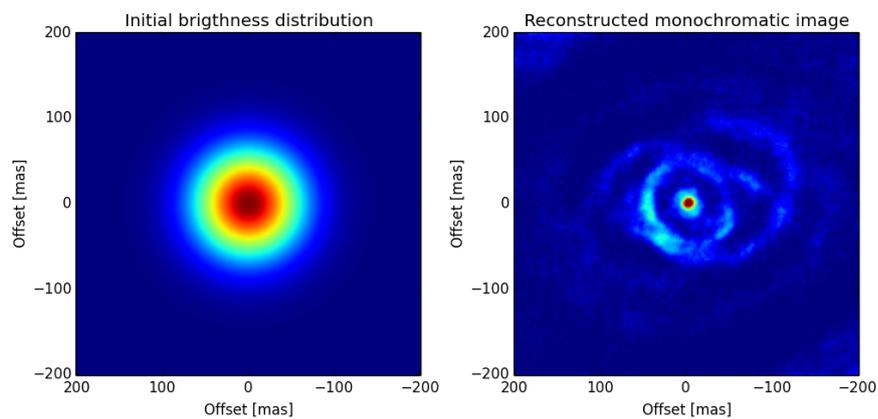

Figure 21: **left:** Initial image used for the monochromatic reconstruction. **right:** Best monochromatic reconstructed image. Notice how in the reconstructed image a rim with an asymmetric brightness distribution is observed around a central source.

set, the `-wavauto` was used. Contrary to the monochromatic reconstruction, here the differential phases were included. The complete command for the polychromatic reconstruction is the following:

```
squeeze 2016dataset2.oifits -w 135 -s 3.0 -la 1000 -l0 1 -ts 3 - fv 0 -i
    initial_monochromatic.fits -novisamp -not3amp -novisphi -wavauto -diffvis -n 150
    -chains 3
```

Figure 22 displays the recovered images channel by channel, as well as the images of the model used to simulate the data. Notice how the asymmetry of the disk is recovered, together with the changes in size and brightness of the central source. However, the elongation of the disk and the internal gap were not recovered in the reconstruction.

# D - THE IMAGE RECONSTRUCTION GRAPHICAL INTERFACE
## D.1 INTRODUCTION
This section describes the graphical user interface that is the central part to operate image reconstruction softwares.

### D.1.1 ARCHITECTURE
It is fine by design to separate the graphical user interfaces from the processing. By this way each blocks are specialized to its limited domain and work by cooperation. The strong advantage to put clear interfaces between each of them help to exchange some part or have choice to compare and run multiple algorithms on same input data and parameters. We then decided to provide a common Graphical User Interface (GUI) that can operates many softwares that respect the interface [6].

The execution architecture of the GUI can handle local executions or remote ones. Remote execution is performed over a standardized execution framework (UWS [7]) which avoid installation of local software to the user. Softwares then are hosted on computing servers. Tests have been performed but first version requires the user to install image reconstruction software on his machine.

## D.2 INSTALLATION AND CONFIGURATION
### D.2.1 REQUIREMENTS
The GUI is packaged as a JavaWebStart [8] application then requires a Java runtime environment. The required command is `javaws`.

Java runs on every operating systems with a great compatibility and efficiency. Users who don't have java on their machine yet, just have to install the java package of their dis-

---

[6] https://github.com/emmt/OI-Imaging-JRA
[7] http://www.ivoa.net/documents/UWS/index.html
[8] https://java.com/en/download/faq/java_webstart.xml

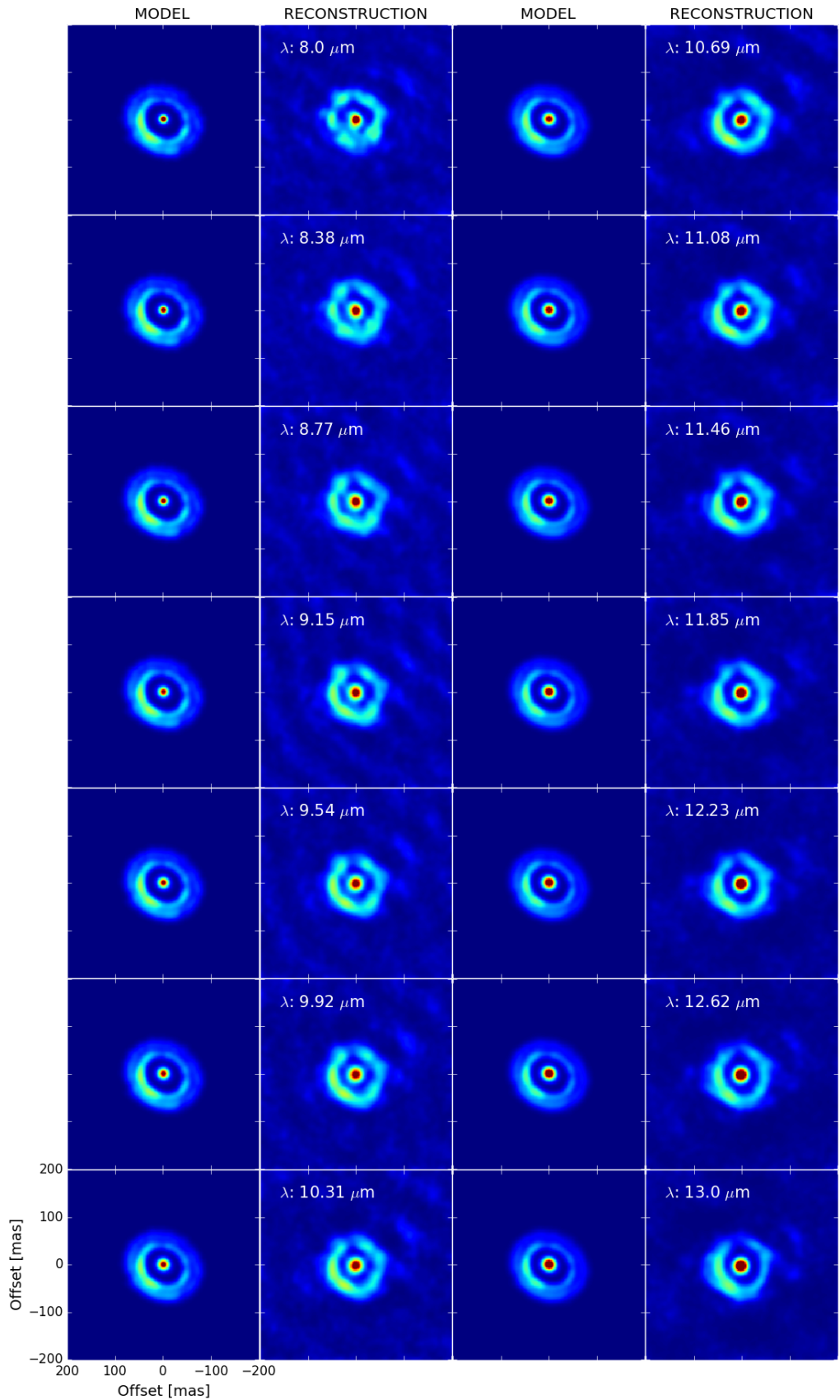

Figure 22: Fourteen reconstructed images across the N-band using the SQUEEZE polychromatic reconstruction. The respective channel is labeled in each frame. The model images (from which the observables were obtained) are also displayed.

tribution or may also visit http://www.java.com. The website provides support in many languages to install the runtime environment.

Image reconstruction softwares must be installed previously and present in the PATH of the GUI. Look at Help Menu / Report Feedback to JMMC ... / Details / System Properties.

Enhancement are planned to help the GUI finding some softwares.

## D.3 BASIC USAGE

The work-flow consists of selecting the input data (`OIFITS`), a starting image (`FITS`) and the image reconstruction software with associated parameters settings. Then the user presses the run button so the algorithm runs until reaching the stopping criterion or maximum number of iterations. Completed runs provide the result data: an image, `OIFITS` tables, and associated output parameters. An execution log is shown and may help the user for any faulty run.

### D.3.1 INPUT DATA

#### D.3.1.1 INTERFEROMETRIC DATA

The GUI load interferometric data through `OIFITS` files. The user can load files from its local disk or may use SAMP protocol to load data from the OiDB portal [9]. `OIFITS V2` is not yet supported by the first version of the GUI.

#### D.3.1.2 IMAGE DATA

Several fits images can be loaded and displayed by the GUI. The selection is performed by the `INIT_IMG` combo box.

#### D.3.1.3 ALGORITHM SETTINGS

A first set of parameters are shown to the user. These parameters come from the interface document.

### D.3.2 OUTPUT RESULTS

#### D.3.2.1 MODEL OF THE DATA

Result `OIFITS` have new column dedicated to model data. A plotting preset has been implemented to simultaneously display the observed data points and modeled ones. Future enhancement already are planned for the shared component developed by JMMC so that the observed and modeled data are stacked on the same plot.

#### D.3.2.2 FINAL IMAGE

As input images, the result image is displayed automatically after each run. The user can compare manually the various result by selecting result set in the dedicated list. Again, some computation ad comparison function are expected in the future versions.

---

[9] http://oidb.jmmc.fr/search

### D.3.2.3 PARAMETERS

A summary of output parameters is display as a basic table. The input data also are displayed to get memory of initial conditions.

### D.3.2.4 EXECUTION LOG

The output of the spawn programs is collected and displayed to the user as free text.

## D.4 APPLICATION MAINTENANCE

The current GUI has been based on the jMCS[10] framework provided by JMMC for its softwares. JMMC will provide support, maintain and develop new functions on this application. Current and future version will be reachable at http://www.jmmc.fr/oimaging. Bug reports, feature requests or documentation typo are submitted to the JMMC's project management system. Some issues already have been filed directly from the GUI at first development phases.

# E - IMAGE RECONSTRUCTION WITH BSMEM

## E.1 INTRODUCTION

The BSMEM (BiSpectrum Maximum Entropy Method) software was first written in 1992 to demonstrate image reconstruction from optical aperture synthesis data. It has been extensively enhanced and tested since then (Baron and Young, 2008). A modified version of BSMEM has been developed as one of several alternative algorithms able to be controlled by the "OImaging" Graphical User Interface presented in C.1. It is this new version of BSMEM that is described here.

The BSMEM algorithm applies a fully Bayesian approach to the inverse problem of finding the most probable image given the evidence, making use of the Maximum Entropy approach to maximize the posterior probability of an image. As of 2017, BSMEM is available free-of-charge for non-commercial use, licensed under a Creative Commons BY-NC-ND licence (see https://creativecommons.org/licenses/by-nc-nd/3.0/). New releases of the software will be made available at:

http://www.mrao.cam.ac.uk/research/optical-interferometry/bsmem-software/

BSMEM uses a trust region method with non-linear conjugate gradient steps to minimize the sum of the log(likelihood) (chi-squared) of the data given the image and a regularization term expressed as the Gull-Skilling entropy:

$$f_{\text{prior}}(\mathbf{x}) = \sum_n \left[ x_n \log(x_n/\bar{x}_n) - x_n + \bar{x}_n \right] \tag{2}$$

with $\bar{\mathbf{x}}$ the *default image*; that is, the one which would be recovered in the absence of any data. The likelihood term used by BSMEM assumes independent Gaussian noise statistics for the amplitude and phase of the measured bispectrum. The optimization engine is

---

[10] https://github.com/JMMC-OpenDev/jMCS

MEMSYS which implements the strategy proposed by Skilling and Bryan (1984) and automatically finds the most likely value of the hyperparameter $\mu$ that weights $f_\text{prior}$ with respect to $f_\text{data}$. Because it doesn't attempt to convert the data into complex visibilities, a strength of BSMEM is that it can handle any kind of data sparsity (such as missing closure phases).

The default image (or model image) $\bar{\mathbf{x}}$ is usually chosen to be a Gaussian, a uniform disk, or a delta-function centered in the field of view, which conveniently fixes the location of the reconstructed object (the bispectra and power spectra being invariant to translation). This type of default image also acts as a support constraint by penalizing the presence of flux far from the center of the image. If a low-resolution image of the target object is available, this can be used instead.

## E.2 INSTALLATION AND CONFIGURATION

These instructions describe local installation of BSMEM on your computer. Installation under Linux or Mac OS X is supported. OImaging can work with a local installation of BSMEM or by running BSMEM on a remote server; if using the latter option local installation is not needed.

First, download the gzipped tar file of the source code[11]. Check the README.md in the tarball for more up-to-date installation instructions.

BSMEM has successfully been built on Linux (Ubuntu 14.04, kernel 4.4.0 and CentOS release 6.6, kernel 2.6.32) and OS X (10.9 Mavericks). Any recent Linux distribution or OS X version should work. Building on Solaris is no longer supported.

### E.2.1 REQUIREMENTS

The following libraries must be installed on your system:

1. PGPLOT 5.2 (http://www.astro.caltech.edu/~tjp/pgplot/): with at least XTERM, PS, and XSERVE drivers activated in the drivers.list file. On Redhat/Fedora Linux install the pgplot-devel package (from rpmfusion); on Debian/Ubuntu install the pgplot5 package. If using MacPorts on OS X install the pgplot port.
2. CFITSIO 3.x (http://heasarc.gsfc.nasa.gov/fitsio/)
   On Redhat/Fedora install the cfitsio-devel package; on Debian/Ubuntu install the libcfitsio3-dev package. If using MacPorts install the cfitsio port.
3. FFTW 3.x (http://www.fftw.org/)
   On Redhat/Fedora install the fftw-devel package; on Debian/Ubuntu install the libfftw3-dev package. If using MacPorts install the fftw-3 port.
4. NFFT 3.x (http://www-user.tu-chemnitz.de/~potts/nfft/)
   Download the source code and build/install using the standard sequence of commands: `./configure`, `make`, `make install`. Alternatively, if using Debian Linux, you can install the libnfft3-dev package.

---

[11] http://www.mrao.cam.ac.uk/research/optical-interferometry/bsmem-software/

5. GLib 2.16 or later (https://developer.gnome.org/glib/)
   On Redhat/Fedora install the glib2-devel package; on Debian/Ubuntu install the libglib2.0-dev package. If using MacPorts install the glib2 port.
6. OIFITSlib 2.2.0 or later (https://github.com/jsy1001/oifitslib/)
7. CMake 2.8.6 or later (https://cmake.org/)

You will also require both C and Fortran 77 compilers to build BSMEM. Recommended compilers are discussed below.

### E.2.1.1 BUILDING PGPLOT FROM SOURCE

If you need to build PGPLOT from source:

Ensure you have an X11 library installed (e.g. libX11-devel and libXt-devel packages on Redhat/Fedora, libX11-dev and libxt-dev on Debian/Ubuntu). Create the makefile using `makemake` (see `install-unix.txt`), then compile. Check that the C wrapper cpgplot is installed. Finally, remember to set up your `PGPLOT_DIR` environment variable to correctly point to the `pgxwin_server`, `rgb.txt`, and `grfont.dat` files.

### E.2.2 RECOMMENDED COMPILERS

BSMEM has been built successfully using gcc, together with g77 or gfortran. On OS X, we recommend the use of gcc 4.5 from MacPorts. Install it and build BSMEM as follows:

- Install MacPorts (https://www.macports.org/)
- Use MacPorts to install gcc45 and gcc_select
- Select this gcc and gfortran:

```
sudo port select --set gcc mp-gcc45
```

- Check that `gcc --version` and `gfortran --version` show version 4.5.4
- Make sure that BSMEM's dependencies are installed (see above)
- Set `PKG_CONFIG_PATH` to include locations of .pc files for libraries not installed through MacPorts, in addition to the MacPorts .pc files directory (normally /opt/local/lib/pkgconfig)
- Build BSMEM using CMake as outlined in the next section, but invoke cmake as `cmake -DCMAKE_C_COMPILER=gcc ..`

### E.2.3 BUILDING BSMEM

CMake is now used for building BSMEM. For those not familiar with CMake, instructions can be found at https://cmake.org/runningcmake/. If you are using a Unix-like operating system the following commands should build the software:

```
cd build
cmake ..
make
```

Finally, use `sudo make install` to install the `bsmem` and `bsmem-ci` executables.

Note that the pkg-config utility is used to locate the GLib and OIFITSLib libraries that bsmem depends on. The search path for pkg-config must be configured to include the locations of the .pc files for those libraries, for example by setting the `PKG_CONFIG_PATH` environment variable.

## E.3 BASIC USAGE

The basic workflow consists of specifying the input data, a starting image for the optimization (this typically also serves as the default image for the entropy function), and the algorithm settings, then starting BSMEM. The algorithm runs iteratively, until either the stopping criterion has been satisfied, or the specified maximum number of iterations has been reached.

### E.3.1 DATA SELECTION

The interferometric data are specified as an OIFITS filename plus data selection parameters that define the subset of the file contents to use. These parameters specify the target object, a wavelength range, and the types of observable to use. BSMEM ignores the `USE_VIS` parameter since the current version cannot use complex visibility data. The non-standard `UV_MAX` parameter allows a maximum radius in the $u-v$ plane to be specified (in waves).

The reconstructed image is assumed to be wavelength-independent, so selecting an appropriate wavelength range from spectrally-dispersed data is a trade-off between minimizing the wavelength-variation in the selected data and maximizing the $u-v$ coverage. If in doubt, select a single wavelength channel.

### E.3.2 INITIAL/DEFAULT IMAGE

If little is known about the object to be reconstructed, or an uninformative prior is desired, a good choice for the default image is usually a circular Gaussian centered in the field of view. With this kind of prior and adequate $u-v$ sampling, getting BSMEM to work usually boils down to choosing the pixel scale, image width, and prior width appropriately. These choices are explored in the examples in §E.4, §E.5 and §E.6.1.

### E.3.3 TOTAL FLUX CONSTRAINT

The image supplied to BSMEM specifies the relative pixel values; the total flux of the initial image is specified by the (currently non-standard) `INITFLUX` parameter. The initial flux should be set to a low value (e.g. 0.01). As BSMEM iterates, the flux in the image model is increased. The flux should be close to unity at convergence. In case there are no measured data at low spatial frequencies, this is enforced by adding an artificial $V^2 = 1$ data point at zero spatial frequency. The `FLUXERR` parameter specifies the error bar on this data point and thus the strength of the "unit flux" constraint. Specifying a `FLUXERR` value

that is too small will significantly delay convergence.

### E.3.4 STOPPING CRITERION

BSMEM is able to estimate the value of the hyperparameter $\mu$ from the data. The `AUTO_WGT` parameter enables or disables this behaviour. If enabled, BSMEM automatically adjusts `RGL_WGT` to obtain a reduced $\chi^2$ close to unity (this is the non-Bayesian "historic chi-squared = N method", which was found to be the most robust of the methods supported by MEMSYS).

If `AUTO_WGT` is disabled, then the behaviour of BSMEM is similar to MiRA. The user must choose an appropriate regularization weight for each imaging problem.

`AUTO_RGL` should be disabled if there is significant intrinsic variation in the selected data, either with wavelength or time (for time-variable targets), as "good" solutions for such cases should have $\chi^2 > 1$. Note that it is not currently possible to filter the OIFITS data by time within OImaging.

### E.3.5 OTHER SETTINGS

The standard algorithm settings are described in D. BSMEM accepts the following non-standard settings (at the time of writing, only FLUXERR can be set in OImaging):

**FLUXERR**  Error on synthetic zero-baseline squared visibility, see §E.3.3

**INITFLUX**  Initial image flux, see §E.3.3

**V2A**  Squared visibility error factor

**T3AMPA**  Triple amplitude error factor

**T3PHIA**  Closure phase error factor

**V2B**  Squared visibility error offset

**T3AMPB**  Triple amplitude error offset

**T3PHIB**  Closure phase error offset

The last six settings can be used to adjust the error bars for the observed data, if you suspect they are systematically incorrect.

### E.3.6 OUTPUT PARAMETERS

The following values can be displayed when BSMEM has finished:

**NITER**  The total number of iterations done in the current program run.

**CHISQ**  The reduced $\chi^2$ at the end of the run.

**ENTROPY**  The entropy value at the end of the run.

**FLUX**  The total image flux at the end of the run.

**RGL_WGT**  The weight of the regularization (determined automatically if the AUTO_WGT setting is true).

## E.4 EXAMPLE 1

For the first example, the model of LkHa 101 used for the 2004 Beauty Contest (Lawson et al., 2004) has been selected (see §F). The following steps are needed to reconstruct an image from this data set, using BSMEM via the OImaging GUI:

1. Prepare an initial image in FITS format (currently this must be done outside of OImaging). The initial image defines the pixel scale and field of view of the reconstructed image.
2. Load the OIFITS file into OImaging. By default, all of the data are used.
3. Load the initial image into OImaging.
4. Adjust the algorithm settings as necessary. The default settings will work for this data set.
5. Click the "Run" button to launch BSMEM. When BSMEM has finished, the reconstructed image will appear in the "Image" tab of the "Data Visualisation" panel (see fig. 23).
6. Examine the "Output parameters" and "Execution log" tabs of the "Data Visualisation" panel to check that the algorithm converged successfully.
7. The "Save image" button may be used to save the reconstructed image.

The initial image used to obtain the reconstruction in fig. 23 was a circular Gaussian with a FWHM of 12 mas, centered in the FOV. The same image was used as the default image for the regularization term (this is the default behavior). The image scale was chosen as 0.25 milliarcseconds per pixel. This is equivalent to 6.8 pixels per fastest fringe $\lambda_{\min}/B_{\max}$ (BSMEM requires a minimum of 6).

When BSMEM starts, the image model has a low flux (given by INITFLUX, default 0.01). As BSMEM iterates, the flux in the image model is increased. fig. 24 shows the squared visibilities corresponding to the image model after 8 iterations, when the model flux is 0.71. At convergence after 29 iterations, the model flux is 1.004 giving a good fit to the shortest-baseline $V^2$ data.

The FLUXERR parameter specifies the error bar on a synthetic $V^2 = 1$ data point which helps enforce normalization of the image to unit flux. Reducing this value from the default 0.1 causes BSMEM to add more flux to the image in later iterations. If there are few real data points at low spatial frequencies, there is little information on where to put this extra flux, leading to spurious results. Hence FLUXERR should be varied with caution.

Use of a centrally-peaked default image is recommended, as this constrains the location of the reconstructed object (squared visibility and triple product data being invariant to translation). In this example, a circular Gaussian of FWHM 12 mas was chosen. Similar results are obtained with a Gaussian FWHM between 9 and 16 mas. Using a

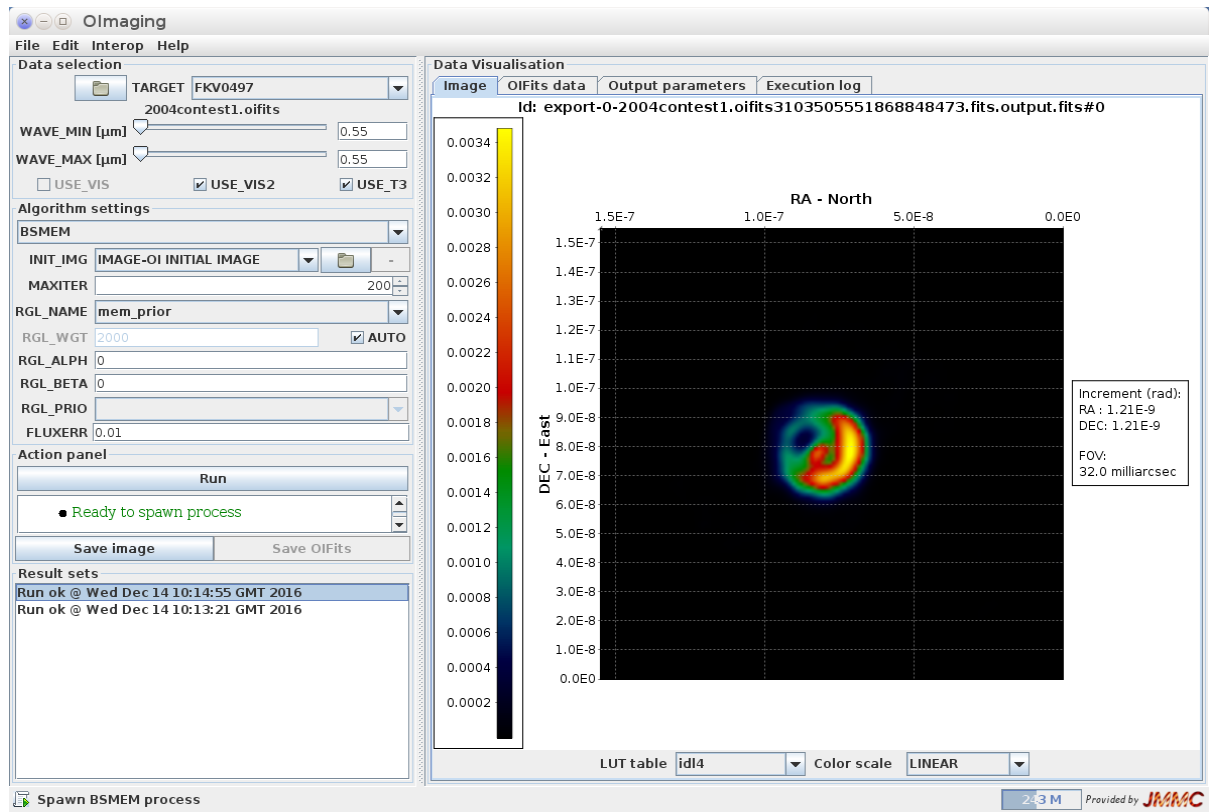

Figure 23: Screenshot of `OImaging` showing input parameters on the left and the final reconstructed image on the right. Note that the color scale is different from fig. 9. The algorithm settings are all at their default values except for `FLUXERR`, which was changed from 0.1 to 0.01.


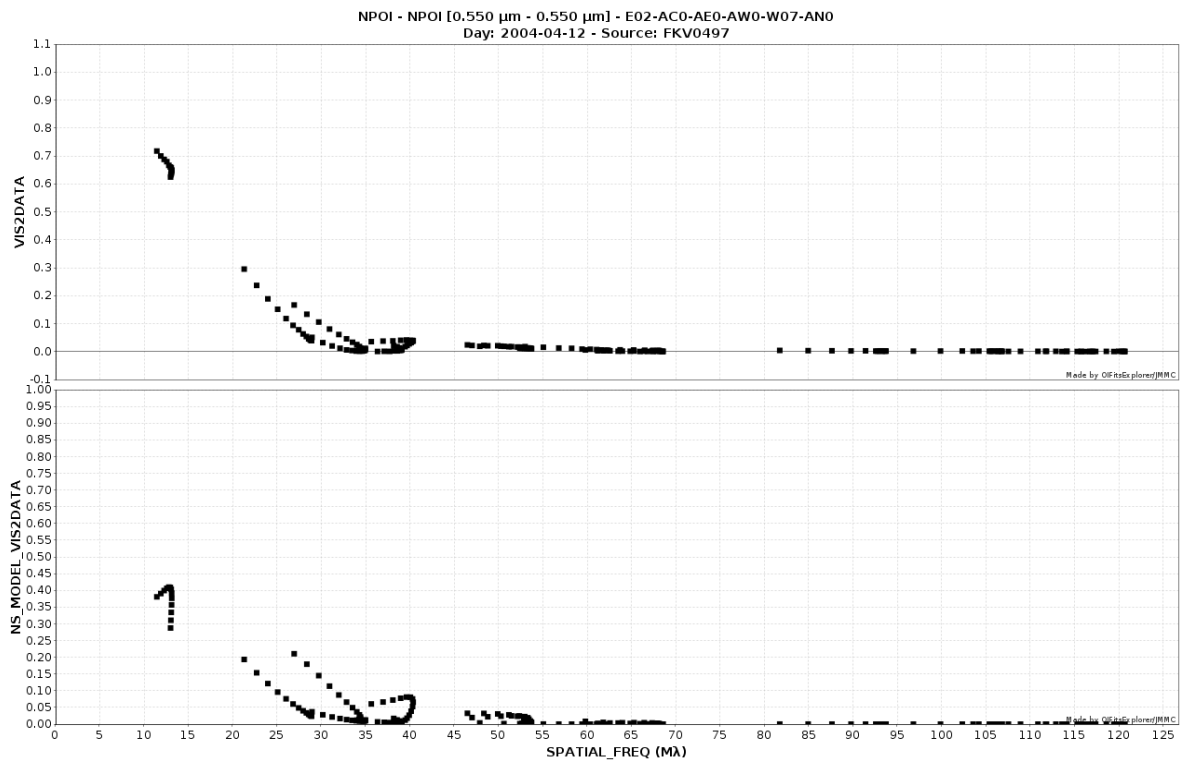

Figure 24: Plot of the input squared visibility data (top) and BSMEM's model of the data (bottom) after 8 iterations.

smaller width allows the reconstruction to converge faster, but increases the risk of missing real flux located far from the FOV center. The suggested approach is to start with a broad default image, infer the true size of the object from the result, then repeat the reconstruction using an appropriately-sized default image. Alternatively, the object size could be determined by fitting a simple model to the short baseline visibilities.

## E.5 EXAMPLE 2

The second example data set is a simulated observation of a proto-planetary disk by VLTI/MATISSE. The synthetic data comprise observations in 14 spectral channels across the N-band (8–13 $\mu$m), using three configurations of the Auxiliary Telescopes. The data set used was part of the 2016 Beauty Contest (Sanchez-Bermudez et al., 2016) and can be found here: www.opticalinterferometry.com

The model includes a dust-enshrouded central star and a disk containing one gap, supposedly carved by a forming planet (Gonzalez et al., 2015). Further details, including plots of the simulated interferometric data, are given in §F.

### E.5.1 GRAY LIMITATIONS

This example illustrates several issues associated with spectrally-dispersed data. Most importantly, BSMEM can only reconstruct a gray image. That is, BSMEM assumes all of the

observed data are consistent with a single truth image, and attempts to find the "simplest" image that is statistically compatible with the data. In general, the user must choose a subset of the available channels in order to optimize the $u-v$ coverage while minimizing the intrinsic variation between channels that BSMEM cannot solve for. An obvious consequence of the gray image limitation is that differential phase data cannot be used.

The $u-v$ coverage of this example data set is adequate to reconstruct each channel independently, given a suitable initial/default image. However, BSMEM's estimates of the regularization weight (hyperparameter) are not very good for this data set, as evidenced by the surprisingly large and rapid variation in the estimated weight as a function of wavelength. The same array configurations were used to simulate all channels, and the integrated spectrum of the object varies smoothly with wavelength, so *a priori* significant variation would not be expected.

A suitable work-around is to set a fixed regularization weight for all spectral channels. As the weight is no longer automatically chosen to give unit $\chi^2$, its important to check the final $\chi^2$ value is acceptable.

In OImaging, each channel must be selected in turn, and a separate reconstruction carried out. Figure 25 shows the settings and result for the longest-wavelength channel. BSMEM achieves convergence for all 14 channels using a regularization weight of 800, yielding $\chi^2$ values between 0.9 and 1.3.

### E.5.2 INITIAL IMAGE

The initial/default image (fig. 26) is the superposition of two circular Gaussian components centered in the FOV: a 90 mas FWHM component containing 99% of the flux and a 5 mas FWHM component containing 1% of the flux. The larger component provides a support constraint for the disk emission, and the smaller component fixes the location of the central star in the middle of the reconstructed image. Successful reconstructions can be obtained without the second image component, but the object position varies significantly with wavelength. The image scale is 2 mas/pixel.

## E.6 ADVANCED USAGE

Sometimes with more challenging data sets, BSMEM won't converge to a good solution when an uninformative initial/default image is used. In such cases, a two-step procedure can often be used to reconstruct an image successfully. The following example illustrates such a method.

### E.6.1 EXAMPLE 3

In this section an example two-step procedure is presented, using simulated data from the 2010 interferometric imaging beauty contest (Malbet et al., 2010).

The contest data simulates an observation of a red supergiant star with VLTI/AMBER.

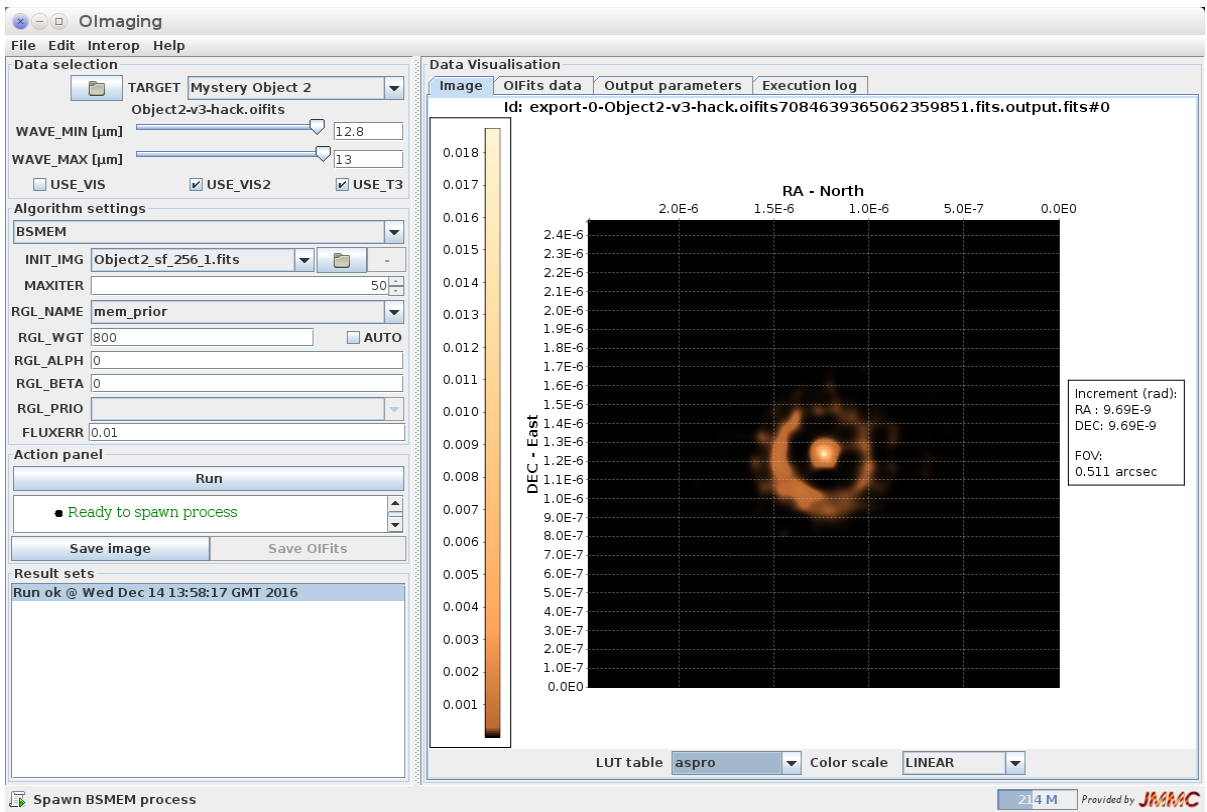

Figure 25: Screenshot of `OImaging` showing input parameters on the left and the final reconstructed image on the right. The longest-wavelength channel has been selected by increasing `WAVE_MIN` to 12.8 $\mu$m. Automatic calculation of `RGL_WGT` has been disabled for the reason explained in the text.

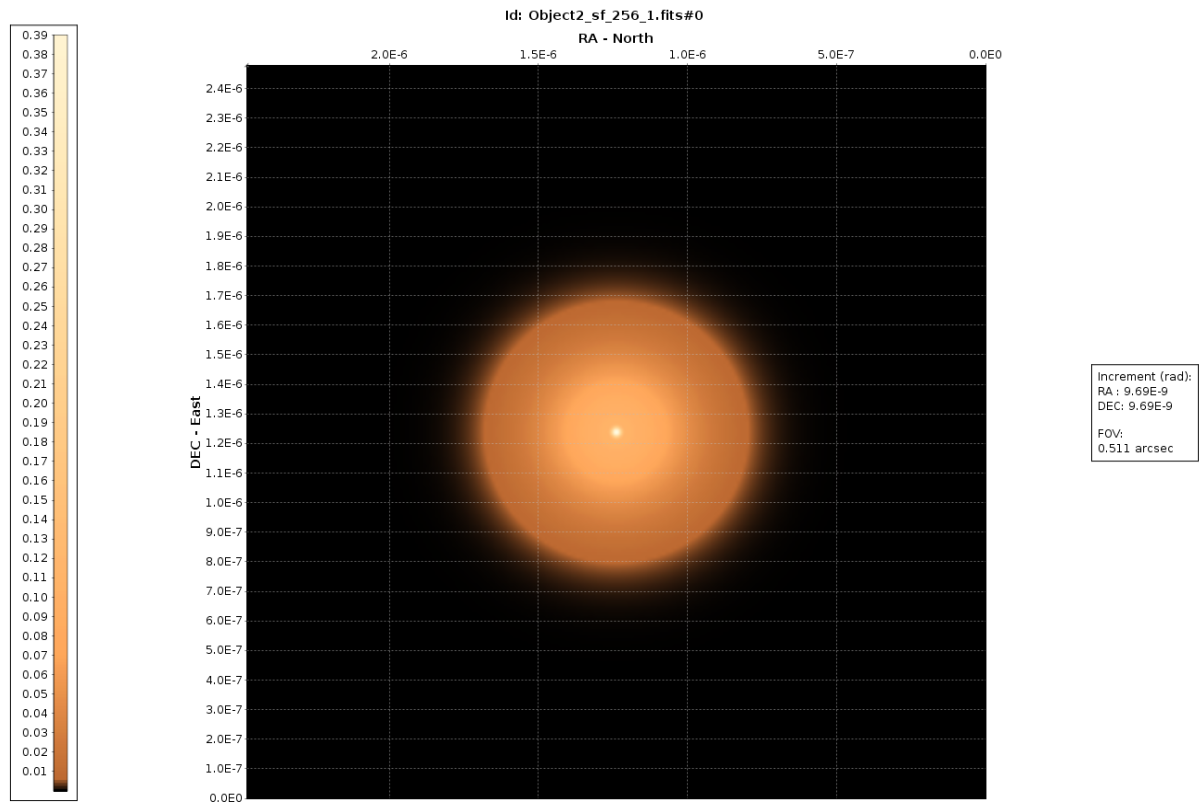

Figure 26: The initial/default image for example 2, comprising a 5 mas FWHM Gaussian superimposed on a 90 mas FWHM Gaussian. See text for explanation.

To add to the challenge, an unresolved companion star, located 8 stellar radii from the primary, was incorporated into the truth image. Two observations of the target object were simulated: a H-band medium spectral resolution (MRH) observation at resolution $R = 1500$ (512 channels), and a simultaneous H- and K-band low spectral resolution observation (LRHK) at $R = 35$ (20 channels).

The contest invited two kinds of submissions:
1. Three "gray" images derived from different subsets of the simulated spectral channels; and
2. A "spectral" image derived from the MRH data, allowing channel-to-channel differences.

The contest data files can be downloaded from the JMMC archive: http://jmmc.fr/oidata/#cat_BeautyContest

### E.6.1.1 BSMEM CONTEST ENTRY

It was found that BSMEM was very slow (runtimes upwards of 24 hours on a standard PC) to converge to a well-fitting (reduced $\chi^2 < 5$) solution when given data from multiple spectral channels and an uninformative prior – probably in large part due to the wavelength-dependence of the object.

Thus a two-step approach was used for the BSMEM contest entry – first finding an im-

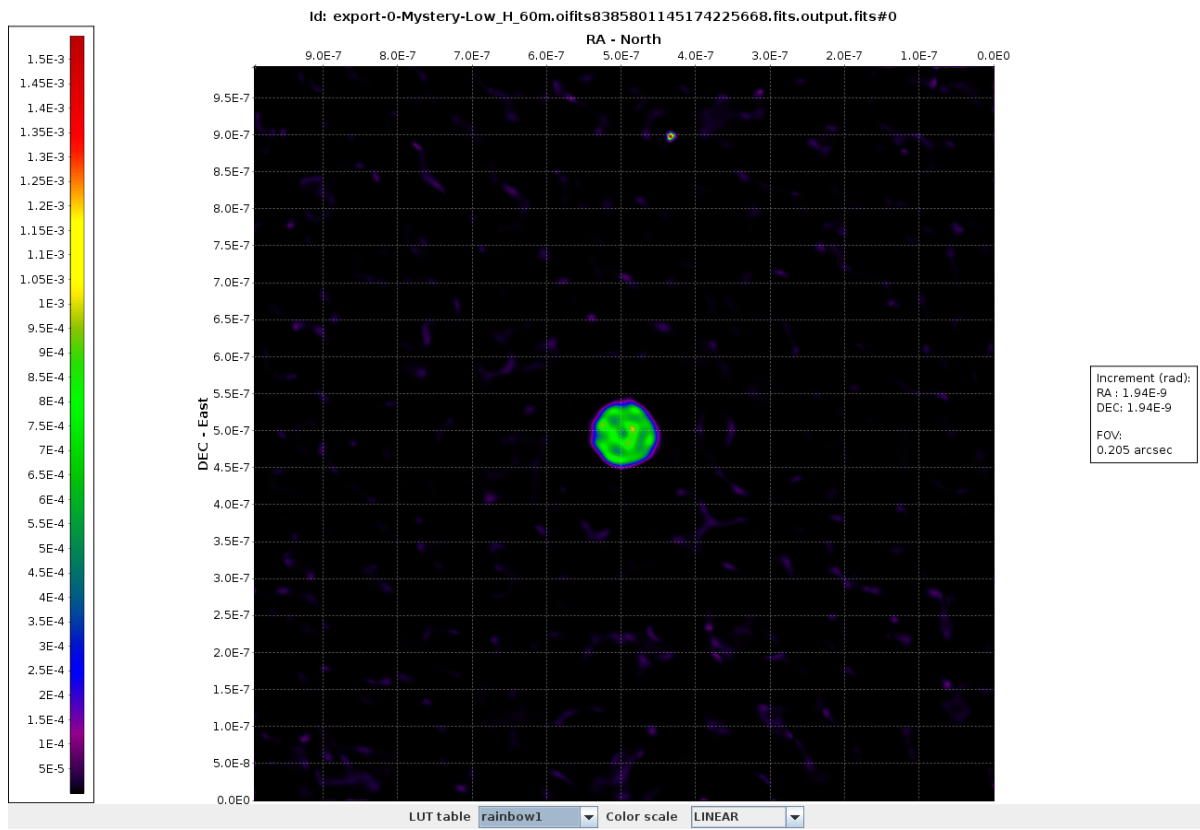

Figure 27: Image reconstructed from a low-spatial-frequency subset ($B < 60$ m) of the LRH data, using a 60 mas Gaussian initial/default image. The companion star is visible towards the top center of the field.

age whose low spatial frequencies were compatible with the data, and then reconstructing each spectral channel separately using this as the default image. The reconstructed images from each spectral channel were then averaged to generate the three gray images specified by the contest organizers.

In particular, it was found that BSMEM did eventually converge using low-spatial-frequency data (baselines up to 60 m) from the first 8 spectral channels of the LRHK data set. The resulting image (fig. 27) was convolved with a Gaussian with FWHM equal to the finest fringe spacing and thresholded to remove features judged to be noise. This spatially-filtered image was used as the initial/default image for the subsequent reconstructions of individual spectral channels, which typically converged after 70 to 250 iterations (of order 30 minutes on a standard PC). To generate the gray submissions, the spectral channels were averaged with uniform weighting, and pixels below a specified threshold in the average image were set to zero. The threshold value was chosen to remove most of the obvious circumstellar artefacts, each of which typically appeared in the one spectral channel only.

When following a two-step process, a pixel scale suitable for the full data set must be used for both steps (the initial image for step 2, derived from the result of step 1, defines the pixel scale for the final reconstruction). A scale of 0.4 mas per pixel gives 6.01

samples per fastest fringe at 1.521 µm and is therefore suitable for this example. With this sampling, a 512 × 512 pixel image is needed to reconstruct both stars.

Since the second-step reconstructions were performed for each spectral channel separately, it was necessary to write a custom script to run `BSMEM` many times, selecting a different subset of the data for each run.

#### E.6.1.2 USING OIMAGING

This section illustrates the use of `OImaging` to perform a similar two-step procedure for a single channel of the 2010 contest LRHK data (`Mystery-Low_HK.oifits`).

For the first step, the high spatial frequencies must be removed from the data. This can be done using the `oifits-filter` program that comes with OIFITSlib. Alternatively, a future `OImaging` release should allow setting of `UV_MAX` to achieve the same result. A 60 mas FWHM circular Gaussian is used as the default image. For this data set, `MAXITER` must be set to a large number as several hundred iterations are needed to converge. The resulting image is shown in fig. 27.

The first reconstructed image is exported and processed outside of `OImaging`. The image is convolved with a Gaussian function, then all pixels below a specified threshold are set to zero. The FWHM of the blurring function is the finest fringe spacing in the filtered data, in this case 5.6 mas. This has the effect of smoothing out all of the stellar surface structure. The threshold value (0.003) is chosen to remove all of the noise in the empty regions of the field.

The blurred and thresholded image is used as the model image for a second `BSMEM` run on the full-resolution data set. For this second step, each spectral channel must be processed separately. fig. 28 shows the result for the first channel.

## F - IMAGE RECONSTRUCTION WITH `MiRA`

### F.1 INTRODUCTION

MiRA, the Multi-aperture Image Reconstruction Algorithm, was developed by Éric Thiébaut,[12] aiming at a versatile package for interferometric image reconstruction, with no restriction on the type of data or regularisation that can be used. The rational is described in detail by Thiébaut (2008, 2013).

MiRA follows a general principle for image reconstruction: it searches for the image $x^+$ (a grid of square pixels) which minimizes a two-term penalty function with respect to the parameters of the image:

$$x^+ = \underset{x \in \Omega}{\arg\min} \left\{ f(x) = f_{\text{data}}(x|d) + \mu f_{\text{prior}}(x) \right\}. \tag{3}$$

The image parameters (e.g., the pixel values) are constrained to the set of possible images $\Omega$ that obey to restrictive conditions, such as *normalisation* (the sum of the pixels is equal to 1) and *positiveness* (all pixel intensities are non-negative).

---

[12]Email: thiebaut@univ-lyon1.fr

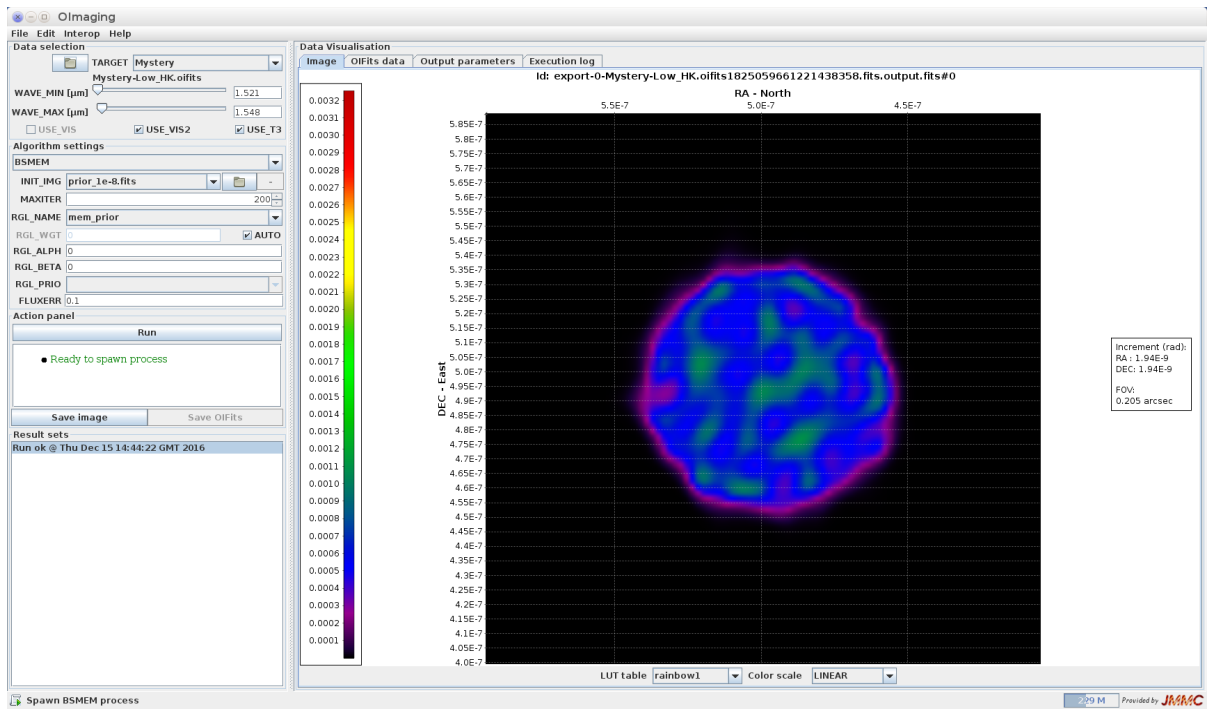

Figure 28: Screenshot of `OImaging` showing input parameters on the left and the reconstructed image (after 72 iterations) on the right. The shortest-wavelength channel has been selected. The image has been zoomed to show the supergiant star surface.

The penalty criterion connects the data and the prior by means of $f_{\text{data}}$ and $f_{\text{prior}}$, usually known respectively as the *likelihood* and *regularisation* terms. While the former measures the discrepancy between the actual data $d$ – e.g., squared visibilities $V^2$, closure phases $\phi_c$, visibility amplitudes $V$ and baseline phases $\phi$ – and their model given the image $x$, the latter is a penalty which enforces additional priors and is required to avoid artefacts. The regularisation term is mandatory because the data alone cannot unambiguously yield an unique image. The *level of regularisation $\mu$* is a positive *hyperparameter*, adjusted to set the relative weight of the a priori information.

The minimisation of eq. (3) is done by a non-linear optimisation algorithm. Since MiRA does not perform a global optimisation, a successful restoration when using closure phases and/or squared visibilities (a non-convex problem) relies on the initial image.

MiRA does not try to recover the phases directly, being thus able to handle any type of observable, and it presents no restrictions on the kind of regularisation to be applied to the reconstruction process.

## F.2 INSTALLATION AND CONFIGURATION

In order to use MiRA, it is necessary to install Yorick (version ≥ 2.2.04), Yeti (version ≥ 6.3.2) and OptimPackLegacy (version ≥ 1.4.0). It is also advisable to install YNFFT, a Yorick plug-in for the *Nonequispaced Fast Fourier Transform*, for faster operations. The

aforementioned packages and software can be retrieved from the following links:
- MiRA: https://github.com/emmt/MiRA
- Yorick: http://dhmunro.github.io/yorick-doc/
- OptimPackLegacy: https://github.com/emmt/OptimPackLegacy
- Yeti: https://github.com/emmt/Yeti
- YNFFT: https://github.com/emmt/ynfft

Yorick, Yeti, OptimPackLegacy, YNFFT and MiRA may be available as packages for your system or can be built from the source files. The source files can be downloaded either as Git repositories or as archive files.

### F.2.1 INSTALLATION OF PRECOMPILED PACKAGES

On Ubuntu based distributions, Yorick and MiRA can be installed using the `apt` terminal command, or by means of the `synaptic` package manager. In case of the former, it is sufficient to write:

```
sudo apt install yorick yorick-mira yorick-ynfft
```

For recent distributions, the previous command will also automatically install `rlwrap`, which is very handy to recall previous commands when running Yorick in interactive mode.

### F.2.2 INSTALLATION OF YORICK FROM SOURCE

#### F.2.2.1 COMPILING AND INSTALLING FROM SOURCE

In case you prefer or need to compile Yorick from scratch, then execute the following steps:[13]

- From https://github.com/dhmunro/yorick get the latest version of Yorick. If you download an archive, unzip/untar it. Example:

    ```
    unzip dhmunro-yorick-y_2_2_04-6-gbc24b71.zip
    ```

    or

    ```
    tar xzvf dhmunro-yorick-y_2_2_04-6-gbc24b71.tar.gz
    ```

    The other possibility is to clone the Git repository of Yorick:

    ```
    git clone git@github.com:dhmunro/yorick.git
    ```

- Go to the unpacked directory or to the Git repository and type:

---

[13]This procedure was tested in `Linux Ubuntu MATE 16.04`, `Mint 17.3 Rosa Mate` and several `Manjaro` and xUbuntu (`Ubuntu, Kubuntu, Ubuntu Gnome`) distributions over the years, both 32 and 64 bits.

```
make config
make clean
make
make check
make intall
```

This will create a subdirectory `./relocate/` in the source tree. The Yorick executable is `./relocate/bin/yorick`. You can move the `./relocate/` directory wherever you want (the name "relocate" is not important), but any changes in the relative locations of the files therein will prevent Yorick from starting correctly. For example:

```
mv relocate /opt/yorick_2.2.04-6__64b
```

You can add the path where the Yorick executable is to your `$PATH` variable, or create a softlink to the Yorick executable from wherever you like, or execute Yorick from a shell script outside its `relocate/` directory.

You may tune the compilation of Yorick during the `make config` step. For instance:

```
make config CC="gcc" CFLAGS="-O2"
make clean
...
```

where optional arguments `CC=...` and `CFLAGS=...` are used to specify the C compiler and optimisation flags respectively. If unspecified, these settings are taken from environment variables[14] (with the same names) or defaults are used. After the `make config` step, you may also edit the file `Make.cfg` to tweak Yorick compilation and installation.

### F.2.2.2 USING A RELOCATABLE VERSION

In case you want to use a previously compiled version of Yorick, just uncompress the `relocatable/` directory to a desired folder (e.g., `/opt`) and add *[Yorick-relocatable-folder ]*`/bin/yorick` to your `$PATH`, or create a soft link to it.

### F.2.3 OPTIMPACKLEGACY, YETI AND MIRA

In order to install any of these packages, just follow the instructions on the corresponding GitHub repositories (see urls in page ).

## F.3 BASIC USAGE

After choosing the regularisation, MiRA needs the following inputs:

---

[14]Environment variables considered by `make config` are: `CC`, `CFLAGS`, `LDFLAGS`, `RANLIB` and `AR`

(i) The data, in the OIFITS standard format, either table versions 1 (Pauls et al., 2005) or 2 (Duvert et al., 2015).
(ii) An optional estimate for the image – important when only power-spectra and bi-spectra are present.
(iii) The pixel size $\delta\theta$ and the size of the reconstructed image.
(iv) The regularisation type and its hyper-parameters, at least $\mu > 0$.
(v) The maximum number of iterations (Gomes et al., 2017).

The initial image is not important if the data is composed of complex visibilities, with both visibility amplitudes and baseline phases, as the problem becomes complex and MiRA yields an unique solution. However, complex visibilities are still scarce and in face of the more common power-spectra and closure phases, the initial image plays an important role in the success of the reconstruction process. The choice of the lateral size $N \times \delta\theta$ of the square image of $N \times N$ pixels is fundamental because it provides a strict constraint on the support of the object and strongly affects the restoration of the image (Gomes et al., 2017). On the one hand, since we are mapping the interferometric data into a discrete grid of pixels, the dimensions of the pixel grid need to be chosen large enough in order to sample all the measured spatial frequencies. On the other hand, $N$ cannot be too large, so as to make the computation practical. According to the Nyquist-Shannon criterion, the pixel size shall sample the maximum angular resolution, but it also shall account for some level of super-resolution introduced by the image reconstruction algorithm. As a rule of thumb, $\delta\theta$ shall be of the order of

$$\delta\theta \lesssim \frac{\lambda}{4\,B_{\max}}, \tag{4}$$

where $B_{\max}$ is the maximum projected baseline length.

MiRA can be used both in the command line interface or inside the `Yorick` environment. The following instructions are highly inspired in the documentation available in the MiRA package (for more details, consult the corresponding GitHub repository, see page 39).

### F.3.1 USING MIRA FROM THE COMMAND LINE

The general syntax for using MiRA from the command line interface is:

```
ymira [OPTIONS] INPUT [...] OUTPUT
```

where `[OPTIONS]` are optional settings, `INPUT [...]` are any number of OIFITS input data files and `OUTPUT` is the name of the output FITS file to save the resulting image. Option `-help` can be used for a short description of all options. Options are prefixed by a single or double hyphen (*e.g.*, `-overwrite` or `--overwrite` are the same) and their order is irrelevant. A double hyphen `--` may be used to mark the end of the options. All input files are read in given order.

### F.3.1.1 DATA SELECTION

If the input data files contain observations for more than one object, the target to consider must be specified as:

```
   ... -target=NAME ...
```

MiRA reconstructs a gray image from the interferometric data. The user may select a wavelength range from the set of observations by means of a combination of keywords. For example, the wavelengths of the selected data can be specified as the end points of the spectral range:

```
   ... -wavemin=MINVAL -wavemax=MAXVAL  ...
```

or as the central wavelength and bandwidth:

```
   ... -effwave=CENTER -effband=WIDTH ...
```

The arguments of these options have units of length. For instance,

```
   ... -wavemin=1.6µm -wavemax=1.8microns ...
```

is the same as:

```
   ... -effwave=1700nm -effband=200nm ...
```

### F.3.1.2 IMAGE PARAMETERS

An initial image for the reconstruction can be provided as

```
   ... -initial=NAME ...
```

where **NAME** is the name of the FITS file with the image to start with or one of **"Dirac"** or **"Random"** to start with a point-like object or a map of random values (the latter being the default). If the reconstruction starts with a random image, option **-seed=NUMBER** can be used to seed the random generator.

By default, the reconstructed image will have the same pixel size and dimensions as the initial image if provided. Otherwise, the pixel size can be specified with the option **-pixelsize=PIXSIZ**, and the image dimensions can be chosen with **-dim=NUMBER** or **-fov=ANGLE**. Arguments **PIXSIZ** and **ANGLE** are in angular units, and **NUMBER** is the number of pixels in each lateral dimension of the image (assumed a square image). For instance,

```
   ... -pixelsize=0.25mas -fov=100mas ...
```

or

```
... -pixelsize=250e-6arcsec -dim=400 ...
```

both yield a 400 × 400 image with a pixel size of 0.25 mas.

F.3.1.3 FOURIER TRANSFORM

The nonequispaced Fourier transform of the pixels can be computed by different methods: `-xform=exact` uses an exact transform, `-xform=nfft` uses a precise approximation by the NFFT algorithm, while `-xform=fft` uses a built-in algorithm which is less precise. The exact transform can be very slow if the image is large and/or if there are many data. The two others exploit an FFT algorithm and are faster. If you have installed Yorick NFFT plug-in, `-xform=nfft` is certainly the method of choice. It is the default method if YNFFT has been installed; otherwise `-xform=exact` is used by default.

F.3.1.4 IMAGE CONSTRAINTS

The total flux of the sought image, and the lower and upper bounds for pixel values can be specified with the options `-normalization=SUM`, `-min=LOWER` and `-max=UPPER` respectively. For instance:

```
... -normalization=1 -min=0 ...
```

is a must for processing OIFITS data.

F.3.1.5 REGULARISATION

Because the algorithm has to deal with sparse interferometric data, the user has also to indicate which regularisation must be applied to interpolate missing data. The regularisation is the kind of prior imposed to the reconstructed image. MiRA accepts any regularisation, but it has several already implemented in the code (see, for e.g., Thiébaut 2013, and `rgl.i`, inside MiRA folder, for details). For all implemented regularisations, the relative strength of the prior (compared to the data) is specified with the option `-mu=`$\mu$, where $\mu \geq 0$.

Of the available regularisations, the following are the most relevant:
- **Edge-preserving smoothness**, which is selected with the options:

  ```
  ... -regul=hyperbolic -mu=μ -tau=τ -eta=η ...
  ```

  where $\tau$ is the edge threshold and $\eta$ the scale of the finite differences to estimate the local gradient of the image. Distinctive scales can be set for different dimensions by providing a list of values to `-eta`, e.g., `-eta=1,1,0.3`. By default, `-eta=1`. Using a very small edge threshold, compared to the norm of the local gradients, mimics the effects of *total variation* (TV) regularisations. Conversely, using a very small edge threshold yields a regularisation comparable to *quadratic smoothness.*

- **Quadratic compactness**, which is selected with the options:

```
... -regul=compactness -mu=μ -gamma=γ ...
```

where $\gamma$ is the full width at half maximum (FWHM) of the prior distribution of light. This parameter has angular units. For instance, `-gamma=15mas`.

### F.3.1.6 CAVEATS

- All units are in SI (i.e., angles are in radians, wavelengths in meters, etc.). A very common error in parameter settings is to use completely out of range values because you assume the wrong units. To make things a easier, some constants are pre-defined by MiRA package, and you can use, for instance, `5*MIRA_MICRON` (instead of $5 \times 10^{-6}$ m), or `3*MIRA_MILLIARCSECOND` (instead of $1.45444 \times 10^{-8}$ rad).
- Positivity is a must, you cannot expect a good image reconstruction (at least with any practical $uv$-coverage) without option `xmin=0` in `mira_solve`. If you use certain regularisations such as *entropy*, you must specify a strictly positive value for `xmin` (choose a very small value, for instance: `1e-50*nrm/npix` where `nrm` is the normalisation level and `npix` the total number of pixels).
- Likewise, `normalization=1` must not be omitted if your data set obeys OIFITS standard (i.e., visibilities are normalised) and has no explicit measurement at frequency $(0,0)$.
- The cost function is highly non-quadratic and may cause difficulties for `OptimPack` to converge. To overcome this, it is sometimes very effective to simply restart `mira_solve` with an initial image provided by a previous reconstruction (possibly after re-centring by `mira_recenter`). For instance,

```
img1 = mira_solve(db, img0, maxeval=500, verb=1, xmin=0.0, normalization=1,
  regul=rgl, mu=1e6);
img2 = mira_recenter(img1);
img2 = mira_solve(db, img2, maxeval=500, verb=1, xmin=0.0, normalization=1,
  regul=rgl, mu=1e6);
img2 = mira_recenter(img2);
img2 = mira_solve(db, img2, maxeval=500, verb=1, xmin=0.0, normalization=1,
  regul=rgl, mu=1e6);
...
```

- It is usually better to work with a purposely too high regularisation level and then lower the value of `mu` as the image reconstruction converges.

### F.3.2 USING MIRA INSIDE YORICK INTERPRETER

The process is usually initiated with the command **mira_new**, such as in

```
db= mira_new("data.oifits");
```

A wavelength included in a set of observations can be selected by means of the keyword `eff_wave`. Example:

```
db= mira_new("data.oifits", eff_wave= 1.617e-6);
```

The image will be restored using data with wavelengths in the range `eff_wave` ± 0.5 × `eff_band` (`eff_band` defaults to 0.1 µm if not input). If `eff_wave` is not specified, MiRA uses the average wavelength computed from the first block of data.

Then, MiRA has to be set-up. The procedure amounts to indicate the dimensions of the restored image, the pixel size and the type of linear transformation for the extrapolation of the observables:

```
mira_config, db, dim= 200, pixelsize= 0.15*MIRA_MILLIARCSECOND, xform= "exact";
```

The keyword `dim`, the number of pixels along the width/height of the restored image, must be indicated in pixels. Alternatively, you can introduce the size of the corresponding FOV, in radians, via the keyword `fov`. MiRA includes several multiples and submultiples of angular units, such as `MIRA_MILLIARCSECOND`, that can be handy when indicating too small or too large values. The `xform` keyword can assume the options `exact` (*exact* Fourier transformsm, the default), `fft` (*Fast* Fourier transforms) or `nfft` (*Nonequispaced Fast* Fourier Transforms).

The regularisation can be indicated with the `rgl_new` command. For example,

```
rgl= rgl_new("compactness");
```

## F.4 EXAMPLES

For the following examples, which illustrate some applications of the MiRA software, it is assumed that all necessary packages are properly installed (see §F.2.3).

### F.4.1 EXAMPLE 1 (INSIDE YORICK INTERPRETER)

Launch Yorick and load MiRA (this shall automatically load Yeti plug-in):

```
include, "mira.i";
```

Load the OIFITS data file (`db` will be our MiRA instance for this data file):

```
db = mira_new("data1.oifits");
```

If there are several spectral channels in the data file, you must choose one with keyword `eff_wave` or choose a spectral range with keywords `eff_wave` and `eff_band`, as described before.

Configure the data instance for image reconstruction parameters:

```
mira_config, db, dim=100, pixelsize=0.5*MIRA_MILLIARCSECOND, xform="exact";
```

Choose a suitable regularisation method:

```
rgl = rgl_new("smoothness");
```

Attempt an image reconstruction (from scratch):

```
dim = mira_get_dim(db);
img0 = array(double, dim, dim);
img0(dim/2, dim/2) = 1.0;
img1 = mira_solve(db, img0, maxeval=500, verb=10, xmin=0.0, normalization=1, regul=
   rgl, mu=1e6);
```

Continue the reconstruction with a re-centred image:

```
img1 = mira_solve(db, mira_recenter(img1), maxeval=500, verb=10, xmin=0.0,
   normalization=1, regul=rgl, mu=1e6);
img1 = mira_solve(db, mira_recenter(img1), maxeval=500, verb=10, xmin=0.0,
   normalization=1, regul=rgl, mu=1e6);
img1 = mira_solve(db, mira_recenter(img1), maxeval=500, verb=10, xmin=0.0,
   normalization=1, regul=rgl, mu=1e6);
```

The resultant image is illustrated in fig. 29.

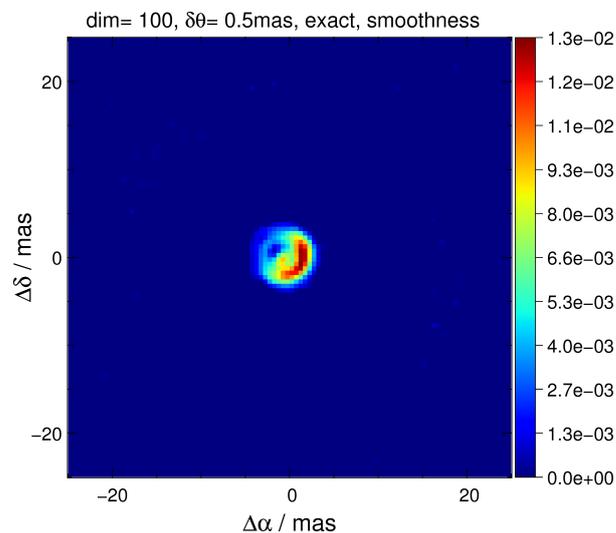

Figure 29: Restored image obtained by MiRA using "smooth" support, lateral width of 100 pixels, and a pixel size of 0.5 mas.

Extract the central part of the image and restart the reconstruction with a higher resolution image and a smaller field of view:

```
cut = (dim + 2)/4;
scale = 4.0;
new_img1 = mira_rescale(img1(1+cut:-cut, 1+cut:-cut), scale=scale);
new_dim = dimsof(new_img1)(2);
new_pixelsize = mira_get_pixelsize(db)/scale;
mira_config, db, dim=new_dim, pixelsize=new_pixelsize;
new_img1 = mira_solve(db, new_img1, maxeval=500, verb=10, xmin=0.0, normalization
   =1, regul=rgl, mu=1e6);
new_img1 = mira_solve(db, mira_recenter(new_img1), maxeval=500, verb=10, xmin=0.0,
   normalization=1, regul=rgl, mu=1e7);
new_img1 = mira_solve(db, mira_recenter(new_img1), maxeval=500, verb=10, xmin=0.0,
   normalization=1, regul=rgl, mu=1e8);
new_img1 = mira_solve(db, mira_recenter(new_img1), maxeval=500, verb=10, xmin=0.0,
   normalization=1, regul=rgl, mu=1e9);
new_img1 = mira_solve(db, mira_recenter(new_img1), maxeval=500, verb=10, xmin=0.0,
   normalization=1, regul=rgl, mu=1e10);
```

Figure 30 illustrates the resultant image.

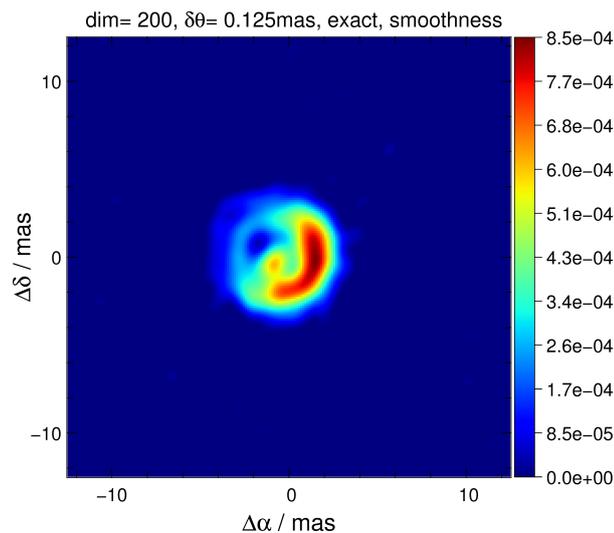

Figure 30: Restored image obtained by MiRA using "smooth" support, lateral width of 200 pixels, and a pixel size of 0.125 mas.

Choose a $\ell_2 - \ell_1$ smoothness:

```
rgl = rgl_new("xsmooth", "cost","cost_l2l1", "threshold",2e-5, "dimlist",dimsof(
   new_img1));
mira_config, db, dim=256, pixelsize=0.1*MIRA_MILLIARCSECOND, xform="fft";
dim = mira_get_dim(db);
r = abs(mira_get_x(db), mira_get_x(db)(-,));
prior = 1.0/(1.0 + (2.0*r/(5.0*MIRA_MILLIARCSECOND))^2);
prior *= 1.0/sum(prior);
```

```
rgl_config, (rgl = rgl_new("quadratic")), "W", linop_new("diagonal",
1.0/prior);
img0 = (prior == max(prior)); img0 *= 1.0/sum(img0);
img1 = mira_solve(db, img0, maxeval=500, verb=1, xmin=0.0, normalization=1, regul=
   rgl, mu=1e4);
img1 = mira_solve(db, img1, maxeval=500, verb=1, xmin=0.0, normalization=1, regul=
   rgl, mu=1e4);
img2 = mira_solve(db, img1, maxeval=500, verb=1, xmin=0.0, normalization=1, regul=
   rgl, mu=5e3);
img2 = mira_solve(db, img2, maxeval=500, verb=1, xmin=0.0, normalization=1, regul=
   rgl, mu=5e3);
img3 = mira_solve(db, img2, maxeval=500, verb=1, xmin=0.0, normalization=1, regul=
   rgl, mu=2e3);
img3 = mira_solve(db, img3, maxeval=500, verb=1, xmin=0.0, normalization=1, regul=
   rgl, mu=2e3);
img4 = mira_solve(db, img3, maxeval=500, verb=1, xmin=0.0, normalization=1, regul=
   rgl, mu=1e3);
img4 = mira_solve(db, img4, maxeval=500, verb=1, xmin=0.0, normalization=1, regul=
   rgl, mu=1e3);
```

## F.5 ADVANCED USAGE

When restoring images in the phase closure case, i.e., when the phase information is solely supplied by closure phases, it may happen that the algorithm stalls in local minima, preventing the image to evolve to the optimal or, at least, to a better solution. *Soft thresholding* can be used in order to make MiRA jump out of local minima. It consists of taking as input for the next step the previous restored image subtracted by a percentage of its maximum, assuring that no negative values are passed. This can be achieved, for instance, with the following code:

```
...
img1= mira_solve(db, img0, maxeval=500, verb=10, xmin=0.0, ...);
img2= mira_solve(db, max(0.0, img1 - 0.05*max(img1))), maxeval= 500, verb=10, ...);
...
```

## F.6 TROUBLESHOOTING

1. When compiling `Yorick` from the source (see §F.2.2.1), make sure you have the package `libx11-dev` installed, or you will get compilation errors.
2. If you have an Intel graphics card and are using the nouveau or Intel driver, `Yorick` might crash after running `MiRA` or when trying to use a small font height in plots. If that is the case, install the 75 and 100 dpi bitmap font packages. For Debian based machines, type in the terminal:

```
sudo apt install xfonts-75dpi xfonts-100dpi
```

Test that everything is OK. You can do it inside `Yorick` with, for example

```
plt, ''Hello, World!", 0.35, 0.65, height=6;
```

or by executing a `mira_solve()` command.

# G - IMAGE RECONSTRUCTION WITH WISARD

## G.1 INTRODUCTION

WISARD stands for "Weak-phase Interferometric Sample Alternating Reconstruction Device". Developed around 2005 at ONERA by S.Meimon, the first version of WISARD recovers a monochromatic ("grey") image from power spectrum and closure phase data alone. In contrast to, e.g., MiRA, WISARD first converts the power spectrum and closure phases measured by optical interferometry into equivalent pseudo complex visibilities (so-called "myopic") data akin to what radio astronomy produces. "Equivalent" here meaning 1) that the phases and their variance are computed from, and are compatible with, the measured phase closures and variances and 2) the amplitudes and variances are computed from, and are compatible with, the measured squared visibilities and variances. A subsequent convexification approximation is performed.

From then on, reconstruction proceeds on the same grounds as for other image-reconstruction programs, by minimizing a compound criterion with two terms, one measuring the fit of the reconstruction given the data, the other regularizing the reconstruction based on a variety of priors (positivity, smoothness, etc).

The following text refers mostly to the envisioned use of WISARD through the OImaging GUI, and heavily borrows from the documentation on the "standalone" version available at http://www.mariotti.fr/wisard_page.htm, which contains all relevant information not only on WISARD but also all references to the scientific publications associated with the concepts behind WISARD and the copyrights. In particular, it is expected from the user of WISARD, directly or through OImaging, that he/she obeys the acknowledgments requirements described in http://www.jmmc.fr/wisard/acknowledgement.

## G.2 INSTALLATION AND CONFIGURATION

The installer can be downloaded from the JMMC website at http://www.mariotti.fr/wisard_page.htm.

Although WISARD consists in a set of IDL procedures, it is not necessary to buy this expensive software: the alternative, open-source, free IDL clone "GDL" is able to run WISARD. GDL is available on several distributions (Debian, FreeBSD but also MacOS and Windows). See http://gnudatalanguage.sourceforge.net/ for details.

We recommend GDL with a version ≥ 0.9.6. `WISARD` uses several procedures and functions from the IDLASTRO library (http://idlastro.gsfc.nasa.gov) and C. Markwardt's library (https://www.physics.wisc.edu/~craigm/idl/). In the following, the terms "IDL" and "GDL" are equivalent.

After unpacking the software (see below), you need to accept the license, which can be accessed under IDL by typing:

```
dummy=wisard(/copyright)
```

### G.2.1 DEPENDENCIES

`WISARD` needs Éric Thiébaut's "OptimPack" library for IDL (`OptimPack_IDL*.so`) and its IDL frontend (`op_*.pro` files) to be installed. For convenience, the OptimPack IDL frontend and several pre-compiled OptimPack libraries for IDL (for Linux, Mac/OS and Solaris, on 32 and 64-bit architectures, and for Windows) are included in the `lib/optimpacklib/` directory of the `WISARD` distribution, with their author's permission. `WISARD` selects the library automatically so this should be transparent. In case of trouble, refer to the complete documentation on the JMMC website.

### G.2.2 UNPACKING THE SOFTWARE

`WISARD` is distributed in a compressed tar archive. Unpacking it is straightforward and will create a `WISARD/` directory under which is the distribution. From the shell prompt, type:

```
tar xzvf Wisard-<version>.tgz
```

## G.3 BASIC USAGE

As for other image-reconstruction programs, `WISARD` will need an input data set (one or more `OIFITS` file — one file with OImaging), a field-of-view value (in milliseconds of arc), and a choice between several regularisations. A maximum value for the number of iterations or, alternatively, a convergence threshold, can be given to limit the number of iterations. Additionally, an input "guess" image can be used as a starting point. In the absence of a guess image, `WISARD` uses the *dirty map* (thresholded to positive values, directly computed as the Fourier transform of the myopic complex visibilities). Other input options are possible, for example to set the minimum number of pixels reconstructed (by default this value is the minimum number needed to fulfill the Shannon-Nyquist criterion).

The choice of the Field of View (`FOV` in the OImaging interface) is critical for the goodness of the `WISARD` reconstruction.

`WISARD` takes all tables in the `OIFITS` file(s) passed and merges them in the minimal set of tables possible: one table per interferometric observable (V2, closure...), array,

object,spectral setup. If the data contains several objects, the one for which image reconstruction should be attempted should be given. Besides there are some constraints on the data (next section).

### G.3.1 CONSTRAINTS ON THE INPUT DATA

`WISARD` quickly evades the problematic of image reconstruction in optical interferometry by using "equivalent" radio-like complex visibilities. By doing so, it poses severe constraints on the observational data:

- `WISARD` needs to correctly associate triplets and visibility baselines. That is, any closure measurement should be associated with 3 existing square visibilities (one on each baseline associated with the triplet). This imposes that related closures and visibilities share the same value of TIME or MJD, a constraint not imposed by the `OIFITS` format and not followed by some instruments. It is possible to ask `WISARD` to allow for small differences in the time-stamps of closures wrt. their related square visibilities by giving a delta time (in seconds) in which two TIME or MJD can be deemed equal. The consequence of this restriction is that `WISARD` may retain fewer observations than present in the data.
- Due to its matrix internal treatment, and pending a complete rewrite (yet unforeseen), `WISARD` cannot mix arrays with different numbers of telescopes. `WISARD` will try to determine the number of telescopes used, based on statistics of the baselines present in the data, and can be corrected if it guesses wrong. In the transformation from closures and squared visibilites to complex visibilities, unknown phases are added, but the number of unknowns decreases rapidly with the number of simultaneous telescopes used: it is in percentage 66% for 3 telescopes, 50% for 4 and 33% for 6. `WISARD` will perform way better when it can use more simultaneous telescopes, even if doing so it discards data obtained with fewer telescopes, which can happen if, e.g., some baseline was "lost" during the observations. In the more desperate cases, one can ask `WISARD` to consider the data as only coming from 3 telescopes: all the closures present in the data will be taken into account, but separately, and at the expense of a worse data reconstruction.
- `WISARD` uses only the `OIFITS` observables `VIS2DATA`, `VIS2ERR` and `T3PHI`, `T3PHIERR`. All other observables are ignored.

In summary: inputting `WISARD` with a set of heterogeneous data observed using various instruments, or using various numbers of baselines, will likely result in `WISARD` using only a "homogeneous" subset of all the data and discard all the rest. This selection process is not usually significant with "modern" `OIFITS` made by recent optical interferometry instruments.

### G.3.2 DATA SELECTION

`WISARD` reconstruct a wavelength-independent ("grey") image. Selecting a range of spectral channels (in the case of multi-channel data) is possible and will permit a piece-wise

reconstruction for objects that change shape significatively along the observed spectral band.

### G.3.3 CHOICE OF REGULARIZATION

Beside the usual constraint of positivity of the reconstruction, `WISARD` proposes a choice of regularizations, mathematical expressions of the somewhat fuzzy preconceived idea that the object to be reconstructed has some kind of smoothness (or piece-wise smoothness, or global smoothness apart from some spikes, etc). In `WISARD`, several regularization terms are available to embody this prior knowledge of the solution.

1. A **smooth solution** is obtained by a quadratic regularization, which uses the object's PSD (Power Spectral Density) as input. This regularization needs the PSD, a 2-D map of size `NP_MIN`×`NP_MIN` containing the PSD for the (quadratic) regularization of the reconstruction. Another parameter, MEAN_O, can be passed to this regularization: it is a 2-D map of size `NP_MIN`×`NP_MIN` containing the MEAN Object to be used for regularization of the reconstruction.
   Using the OImaging interface, this is triggered by the `PSD` choice.
2. A **piecewise smooth prior** is obtained by a linear-quadratic (so-called L1L2) regularization. This prior is called *edge-preserving* as it allows sharp edges in the object if the data is compatible with them, contrarily to a quadratic regularization. Associated parameters are DELTA and SCALE:
   - DELTA is a scalar factor for L1-L2 regularization, used to set the threshold between quadratic (L2) and linear (L1) regularization.
   - SCALE is a scalar factor for L1-L2 regularization, which should be of the order of the RMS object's gradient value. More details are available in the JMMC documentation.
   
   Using the OImaging interface, this is triggered by the `L1L2` choice.
3. A **variant** of this regularization, designed to grant the solution with some smoothness while allowing spikes: this pixel-independent (or white) L1L2 regularization is called *spike-preserving*. Associated parameters are DELTA and SCALE as above, except that SCALE should be here of the order of the average object value.
   Using the OImaging interface, this is triggered by the `L1L2WHITE` choice.
4. The **Total Variation** (*totvar*) regularization, which tends to avoid having many small variations in the image but does not object to have a small number of large variations. This preserve extended shapes (making them more uniform than they probably are) and does not remove isolated spikes (e.g., stars). *totvar* is a simple and quite successful regularization for astronomical image restoration.
   Using the OImaging interface, this is triggered by the `TOTVAR` choice.
5. The **soft_support** regularization uses an additional image (map) and forces the reconstructed image towards the shape of this *support* (e.g., a star uniform disk). Since the interchange format for image reconstruction algorithms does not implement this case, this regularization will probably be absent from OImaging. As-

sociated parameters are MU_SUPPORT, FHM and MEAN_O. Refer to the JMMC documentation for more information on this regularization.

In case of implementation in the OImaging interface, this would be triggered by the `SOFT_SUPPORT` choice.

## G.4 STOPPING CRITERION

The stopping criterion occurs after either NBITER iterations or when the THRESHOLD has been attained:

- NBITER is the maximum number of iterations for the reconstruction, 500 by default. For a better control of the reconstruction, one should rather use `THRESHOLD` below and not lower this value.
- THRESHOLD is the convergence threshold to be used as a stopping criterion for the iterations. It is by default set to the machine precision in simple precision (around $10^{-7}$), even if computations are done in double precision. For a (rather) quick-look result, set to a smaller value, e.g., $10^{-6}$.

## G.5 OUTPUT PARAMETERS

`WISARD` in its standalone version outputs at the end of a minimization cycle the reconstructed image and an IDL structure, AUX_OUTPUT, described in the file wisard.pro of the distribution and containing all relevant information. The `wisardgui` procedure used in conjunction with the OImaging GUI reads and writes `OIFITS` files according to the interoperability norm described in the Interface to Image Reconstruction document (https://github.com/emmt/OI-Imaging-JRA/blob/master/doc/interface/OI-Interface.pdf)

### G.5.1 GALLERY

The following figures are examples of what can be expected from `WISARD` depending from the regularization used (and the peculiarities of the object-to-be reconstructed).

## H - CONCLUSIONS

- Restoring milliarcsecond resolution interferometric images in the infrared will represent a key facility for the new generation of optical interferometers. During the last decade, the community has gain experience by developing a number of different softwares and innitiatives (e.g., the Interferometric Imaging Beauty Contest). The packages, here described, are the result of an extensive collaborative effort and represent a reliable basis to properly recover images from interferometric data.
- The current understanding of the image reconstruction problem together with today's development of software allow the user to perform both mono-chromatic and poly-chromatic image reconstructions of simulated and real interferomet-

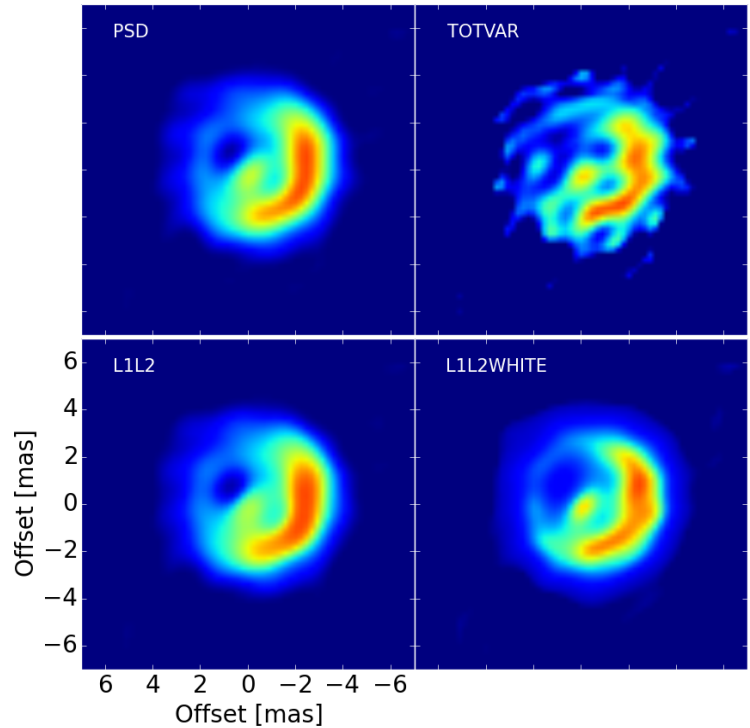

Figure 31: Four reconstructed images of the same object as in previous chapters, stopping criteria is $10^{-6}$, from left to right and top to bottom: PSD, TOTVAR, L1L2 and L1L2WHITE. L1L2WHITE give the best results in this case (smooth object wit a spike at the center), and TOTVAR the worst.

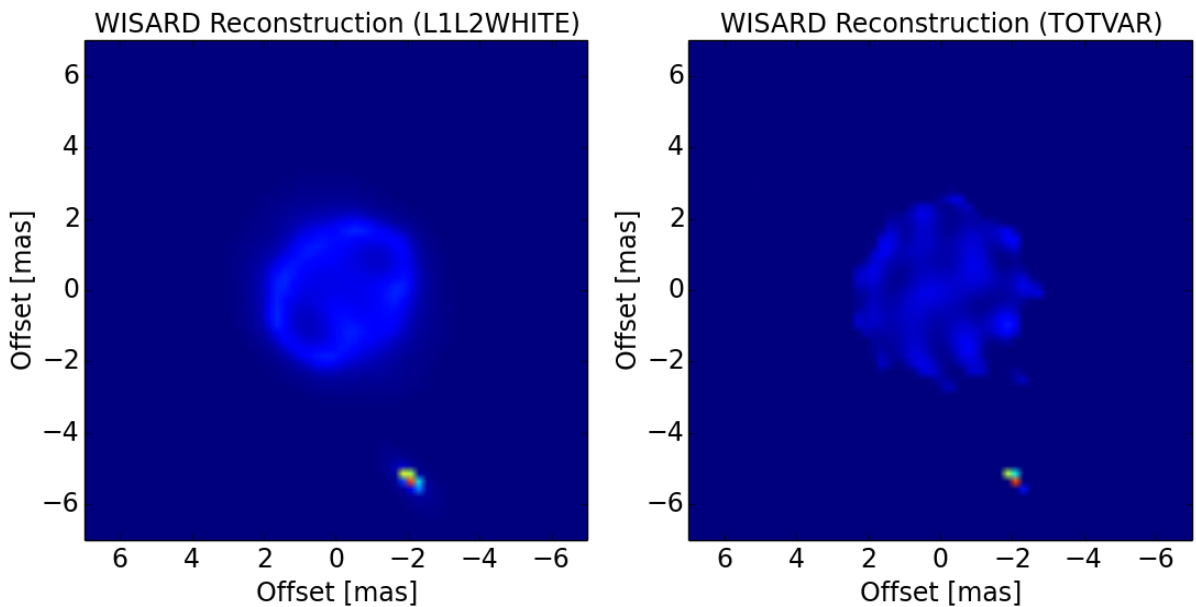

Figure 32: Two reconstructed images of a binary: a bright unresolved source near to the large uniform disk of a supergiant. The scaling of the image is logarithmic as to show the disk of the supergiant star. Left: TOTVAR, right L1L2WHITE. TOTVAR is closer in this case to the flatness of the supergiant surface and the compactness of the unresolved star.

ric data. However, algorithms to perform full poly-chromatic reconstructions are still under development and future improvements of the existing software should move towards this field.
- Testing the image capabilities of the different imaging algorithms have been essential to have a full description of them. This agrees with the main goal of this report, which consists in providing a simple an homogeneous view of image reconstruction in optical interferometry to the community, by having complete cookbooks of the different packages as well as a dedicated GUI to use them.
- The better we understand the requirements to achieve a science-grade images from interferometric observations, the easier will be to provide tools and procedures to the community to make more accessible the use of the current techniques. This is a task that should be addresses in the coming years as part of an effort to broaden and engage the field with more members of the international community.



\end{appendix}

\end{document}